\documentclass[a4paper,usenatbib]{mnras}

\usepackage{newtxtext,newtxmath}
\usepackage{makecell}

\usepackage[T1]{fontenc}
\usepackage{ae,aecompl}

\usepackage{myaasmacros}

\usepackage{graphicx}	
\usepackage{mathrsfs}
\usepackage{hyperref}
\usepackage{pifont}
\usepackage{color}
\usepackage{multirow}

\usepackage{xspace}

\newcommand{\sigmaSFR}{$\Sigma_{\mathrm{SFR}}$\xspace}
\newcommand{\sigmaGas}{$\Sigma_{\mathrm{gas}}$\xspace}

\usepackage{abbrevs}

\begingroup
\catcode`\_=\active
\gdef_#1{\ensuremath{\sb{\mathrm{#1}}}}
\endgroup
\mathcode`\_=\string"8000
\catcode`\_=12

\newcommand{\Msolar}{M$_{\odot}$\xspace}
\newcommand{\Msun}{\ensuremath{\mathrm{M}_\odot }\xspace}

\newcommand{\SATIN}{\textsc{satin}}
\newcommand{\satin}{\textsc{satin}}
\newcommand{\satinI}{\textsc{satin1}}
\newcommand{\SNe}{\textit{SNe}}
\newcommand{\nSNe}{\textit{nSNe}}

\definecolor{darkgreen}{rgb}{0.0,0.7,0.0}

\newcommand{\Ramses}{{\textsc{ramses}}}

\newcommand{\orcid}[1]{\href{https://orcid.org/#1}{\includegraphics[width=10pt]{orcid}}}

\makeatletter
\newcommand*\bigcdot{\mathpalette\bigcdot@{.5}}
\newcommand*\bigcdot@[2]{\mathbin{\vcenter{\hbox{\scalebox{#2}{$\m@th#1\bullet$}}}}}
\makeatother

\makeatletter
\renewcommand\maybe@space@{%
	\maybe@ictrue 
	\expandafter   \@tfor
	\expandafter \reserved@a
	\expandafter :%
	\expandafter =%
	\nospacelist
	\do \t@st@ic
	\ifmaybe@ic 
	\space
	\fi
}
\makeatother

\graphicspath{{Figures/}{Figures/maps/}{Figures/massFlow/}{Pictures/}{/Users/rebekka/SATIN/04_10pc_InitSNe/02_mw_sne_noeq}{/Users/rebekka/SATIN/04_10pc_InitSNe/02_mw_sne_noeq/new_plots}{/Users/rebekka/SATIN/04_10pc_InitSNe/02_mw_sne_noeq/new_plots/KS}}

\newcommand\plotone[1]{%
 \centering
 \leavevmode
 \includegraphics[width={\columnwidth}]{#1}%
}%
\newcommand\plotoneF[1]{%
 \centering
 \leavevmode
 \includegraphics[width={\textwidth}]{#1}%
}%
\newcommand\plottwoH[2]{{%
 \centering
 \leavevmode
 \columnwidth=.5\textwidth
 \includegraphics[width={\columnwidth}]{#1}%
 \includegraphics[width={\columnwidth}]{#2}%
}}%

\title[Turbulent ISM in MW simulations with SNe feedback]{The SATIN project I: Turbulent multi-phase ISM in Milky Way simulations with SNe feedback from stellar clusters}
\author[R. Bieri  et al. ]
{Rebekka Bieri$^{1,2}$\thanks{Corresponding author: Rebekka Bieri (bieri@mpa-garching.mpg.de)}, 
Thorsten Naab$^{1}$, 
Sam Geen$^{3}$, 
Jonathan P. Coles$^{4}$, \newauthor
R\"udiger Pakmor$^{1}$,
and Stefanie Walch$^{5}$ \\
$^{1}$ Max-Planck-Institute for Astrophysics, Karl-Schwartzschild-Strasse 1, 85748 Garching, Germany,\\
$^{2}$ Center for Space and Habitability, University of Bern, Bern, Switzerland \\
$^{3}$ Anton Pannekoek Institute for Astronomy, Universiteit van Amsterdam, Science Park 904, 1098 XH Amsterdam, The Netherlands\\
$^{4}$ Swiss National Supercomputing Centre (CSCS), Lugano, Switzerland \\
$^{5}$ Physikalisches Institut der Universit\"at zu K\"oln, Z\"ulpicher Strasse 77, 50937 K\"oln, Germany\\}
\date{\today}

\begin{document}
\label{firstpage}
\pagerange{\pageref{firstpage}--\pageref{lastpage}}
\maketitle

\begin{abstract}
We introduce the star formation and Supernova (SN) feedback model of the \satin{} (\textbf{S}imulating \textbf{A}GNs \textbf{T}hrough \textbf{I}SM with \textbf{N}on-Equilibrium Effects) project to simulate the evolution of the star forming multi-phase interstellar medium (ISM) of entire disk galaxies. This galaxy-wide implementation of a successful ISM feedback model naturally covers an order of magnitude in gas surface density, shear and radial motions. It is implemented in the adaptive mesh refinement code \Ramses{} at a peak resolution of 9 pc. New stars are represented by star cluster (sink) particles with individual SN delay times for massive stars. With SN feedback, cooling and gravity, the galactic ISM develops a realistic three-phase structure. The star formation rates naturally follow observed scaling relations for the local Milky Way gas surface density. SNe drive additional turbulence in the warm (300 K < T < 10$^4$ K) gas and increase the kinetic energy of the cold gas, cooling out of the warm phase. The majority of the gas leaving the galactic ISM is warm and hot with mass loading factors of $3 \le \eta \le 10$. While the hot gas is leaving the system, the warm and cold gas falls back onto the disc in a galactic fountain flow.
\end{abstract}




\section{Introduction}
\label{sec:intro} 
Understanding the formation of galaxies within their cosmological context poses
significant challenges owing to the large dynamical range in space and time as
well as the wide range of physical mechanisms involved. In the current paradigm
of galaxy formation, dark matter (DM) is gravitationally dominant on large
scales, and is, therefore, crucial to the understanding of structure formation
within the universe. Baryons follow the DM into gravitationally bound structures,
the halos, where they cool, which decreases their pressure and dissipates angular momentum \citep{Binney1977,
Rees+Ostriker1977, Silk1977, White+Rees1978}. Eventually they form galaxies in
their centre.  Further cooling and fragmentation leads to the formation of
giant molecular clouds (GMCs) where stars form
\citep{McKee+Ostriker2007}.  Feedback processes and outflows are needed to
prevent stars from forming too efficiently compared to observations
\citep[e.g.][]{White+Rees1978}. For low-mass galaxies, such effective feedback
is thought to come from massive stars  \citep{Dekel+Silk1986, Efstathiou2000},
whereas for high-mass galaxies,  active
galactic nuclei (AGN) feedback from supermassive black holes (SMBH) are likely the dominant source of feedback \citep[e.g.][]{Silk+Rees1998, Benson+2003}.

It has been shown that feedback from massive stars, in the form of radiation,
momentum, and thermal energy, can deplete cold-gas reservoirs, regulate star formation, and
launch galactic winds in low-mass galaxies. Moreover, it is important in setting
the phase structure and porosity of the multi-phase ISM
\citep[e.g.][]{Stinson+2006, Sales+2010, Dubois+Teyssier2008, Ostriker+Shetty2011, Hopkins+2014, Kimm+Cen2014, Marinacci+2014, Dubois+2014, Vogelsberger2014, Gatto+2015, Murante+2015, Wang+2015, Schaye+2015, Martizzi+2016}. For a detailed review on this see \cite{Somerville+Dave2015} and 
\cite{Naab+Ostriker2017}.
There are various different stellar feedback channels.  Radiation from massive stars
can act directly to suppress star formation by ionising and heating
their immediate environment. Radiation feedback can additionally strengthen galactic
outflows and thus help to lower the total gas content available to form stars
\citep[see e.g.][for some recent simulations]{Walch+2012,Dale+2014,Geen+2015b,Peters+2017,Kim+Ostriker2018,Kim+2019}. Stellar winds of massive stars have been found to disperse their host cloud and
suppress gas accretion onto the newly formed stellar cluster with various levels of efficiency \citep[see e.g.][for most recent simulations]{Lancaster2012, Dale+2014,Mackey+2015,Gatto2017,Haid+2018,Geen+2021, Guszejnov2022}. 
At the end of their lifetime many massive stars explode as SNe, where an over-pressurised
gas bubble expands into, sweeps up, and accelerates ambient material. Depending on the larger scale environment they can
eject gas from the galaxy and even the halo \citep[see e.g.][]{Martizzi+2015,Martizzi+2016,Kim+Ostriker2015, Walch+2015, Walch+Naab2015, Haid+2016, Li+2017, Rathjen+2021}. An additional pressure component that can
drive gas out of the ISM can come from cosmic rays that are generated in the
SN blown shocks and interact with the magnetic field
\citep{Dorfi+Breitschwerdt2012, Simpson+2016, Girichidis+2016, Girichidis+2018, Rathjen+2021, Hanasz+2021}. Other sources of feedback such as the impact of stellar jets
\citep{Verliat+2022, Nakamura+Li2007, Wang+2010, Guszejnov+2021}, high-energy
photons from X-ray binaries \citep[e.g.][]{Kannan+2016}, as well as runaway
stars \citep[e.g.][]{Kimm+Cen2014, Andersson+2021, Steinwandel2022} are still a
relatively new subject of exploration. These additional stellar sources of
feedback are also found to have an impact on star formation as well as outflow properties. However, their respective relevance compared to other stellar feedback
channels may be smaller and is still the subject of discussion. Recent simulations include ever more stellar feedback channels in order to better understand their
relative importance and the details of their interaction with each other \citep{Dale+2014, Geen+2015, Haid+2018, Geen+2021, Rathjen+2021, Grudic+2022}. Such simulations
are increasingly important and reveal how complex it is to properly understand their relative effect on the surrounding gas, star formation, and outflow properties. 

For more massive galaxies, stellar feedback is, however, much less effective in
regulating star formation because of their deeper potential wells that make it
harder for the gas to escape. A more powerful source of feedback for these massive galaxies can be provided by AGN feedback from supermassive black holes (SMBHs)
which are thought to be ubiquitous at the center of massive galaxies
\citep[e.g.,][]{Magorrian1998, Hu2008, Kormendy+2011}.  Rapid gas accretion
onto black holes leads to an energy release capable of driving outflows that
regulate star formation and the baryonic content of galaxies
\citep{Silk+Rees1998}. This, in turn, limits their own growth (leading to efficient
self-regulation of the BH growth) as well as the gas content of the surrounding galaxy
\citep{Kormendy+Ho2013}. Although the general picture of black holes exerting 
feedback on their host galaxies is very attractive, the details remain
vague, as the exact coupling between the AGN and the ISM of the host galaxy is
still poorly understood. High-resolution simulations of different AGN feedback channels
(jets, mechanically and radiation driven winds) have shown that including a
multi-phase gas structure results in different interactions with the ISM
compared to simulations with a homogeneous setup
\citep[e.g.,][]{Bicknell+Sutherland+2000, Sutherland+Bicknell2007,
Antonuccio-Delogu+Silk2010, Wagner+Bicknell2011, Gaibler+2012, Bieri+2017}.
Efforts in understanding the coupling efficiency of AGN feedback with the
surrounding gas must therefore include a multi-phase model of the ISM
structure. 

The multi-phase ISM is composed of three different phases coexisting and
interacting with each other: a \textit{cold phase} consisting of atomic and
molecular gas ($T < 10^4~$K); a \textit{warm phase} ($T \sim 10^4$~K) composed of ionised gas
and largely neutral gas; and a meta-stable \textit{hot phase} (temperatures
exceeding $T = 10^5$~K) of ionised gas produced predominantly by mechanical
energy input from supernovae \citep{Cox+Smith1974, McKee+Ostriker1977,
Ferriere2001, Klessen+Glover2016}.  The ionised gas makes up most of the mass within the MW galaxy, followed by the cold and high density molecular gas \citep[see][for a review]{Saintonge+Catinella2022}. On the other hand, 
most of the volume is occupied by the warm neutral and ionised gas
\citep[e.g.][]{Kalberla+Kerp2009}. 

Turbulence is important for the structure of the ISM, as revealed by
observations and simulations \citep[see review by e.g.][]{MacLow+Klessen2004,
Elmegreen+Scalo2004}. It is an important ingredient for star formation
\citep{McKee+Ostriker2007} and affects the rate at which stars are formed
\citep{Federrath+Klessen2012, Padoan+2014} as well as influencing the global and
local stability properties of galaxies \citep{Romeo+2010, Hoffmann+Romeo2012,
Romeo+Agertz2014, Agertz+2015}. Because turbulence dissipates on small scales,
a mechanism of driving at large scales is needed. The main mechanisms that
maintain turbulence within the ISM are, however, still not clear. There are
several candidates capable of driving turbulence in the ISM: stellar feedback
in the form of supernovae, jets, winds and ionising radiation
\citep[e.g.][]{deAvillez+Breitschwerdt2005, Joung+MacLow2006, Kim+2013,
Iffrig+Hennebelle2015, Girichidis+2016, Padoan+2016}; gravitational
instabilities coupled with galactic rotation \citep{Gammie+1991,
Piontek+Ostriker2004, Wada2008, Bournaud+2010, Renaud+2012,
Krumholz+Burkhart2016, Meidt+2018, Nusser+Silk2022};
Magneto-Rotational-Instability \citep{Balbus+Hawley1991}; as well as shear between 
a cold disc with respect to the hot halo \citep[see discussion in Sec. 3 of][]{Pfrommer+2022}. A multi-phase model
of the ISM gas structure must, therefore, go jointly with the development and
understanding of a stellar feedback model that helps to regulate the (energetic)
structure of the ISM. Additionally, the model of the galaxy ideally also
includes the self-gravity of the gas as well as cooling channels that all
influence the motion and evolution of the gas within the galaxy. 

In recent years, there has been great deal of improvement in numerical simulations
of the ISM and the interaction of the multi-phase gas with various different
stellar feedback channels as well as additional physical ingredients 
\citep[see e.g.][]{Kim+Ostriker2015, Martizzi+2015, Walch+2015, Geen+2015b, Peters+2017, Rathjen+2021}. Such simulations reached ever higher resolutions to understand the
various physical influences within a molecular cloud (MC) as well as their immediate surroundings.
However, as a consequence of their detailed study of the different physical
processes the simulations are performed in idealised setups and rarely include
the larger galactic or even cosmological environment. Recent exceptions are
dwarf galaxy simulations where the total gas mass is small enough to allow for
a more detailed study \citep[e.g.][]{Hu+2016,Hu+2017,Emerick+2019,Lahen+2020,Rey+2020,Gutcke+2022}.

Unlike detailed studies of dwarf galaxies or portions of the ISM, larger-scale simulations 
typically have to rely on sub-grid
models to follow star formation, stellar feedback, and AGN feedback processes.
Due to computational limitations, the internal structure of the multi-phase ISM is, in those simulations, at best marginally resolved \citep[e.g.][]{DiMatteo+Springel+2005, DiMatteo+2008, Sijacki+2007, BoothSchaye2009, duboisetal2012, Hopkins+2014, Vogelsberger2014, Kimm+2015, Schaye+2015, Pillepich+2018}. The sub-grid models used are not
expected to generally produce the same internal structure within a galaxy as those found
in resolved high-resolution feedback simulations. Despite this, larger-scale simulations 
remain valuable because they place galaxies in a more realistic environment, where
major and minor mergers, gas clump capture, as well as cold filamentary
accretion affect the evolution of these galaxies, likely in a way that is not
predicted by idealised simulations. Despite the simplicity of their
modelling, these large-scale studies have highlighted the capacity of stellar and AGN
feedback to regulate the star formation process as well as gas content in small
and massive galaxies, respectively. 

The main limitation of these large-scale simulations is numerical resolution as 
much higher resolution is required to attempt to model the ISM
\citep[see e.g.][]{Kim+Ostriker2015, Martizzi+2015, Walch+2015, Geen+2015b, Peters+2017, Rathjen+2021}. Simulations of MW-like galaxies that start to bridge the gap between such large-scale models and more
detailed simulations of the ISM and individual MCs have only recently started
to become numerically feasible \citep{Agertz+2013, Rosdahl+2015, Grisdale+2017, Hopkins+2018, Martizzi+2019, Marinacci+2019, Tress+2020}.  However, such simulations have not (yet) studied the 
interaction of AGN feedback with the surrounding
multi-phase ISM. But in order to properly quantify the role of AGN in the
evolution of (massive) galaxies we need more theoretical work involving
realistic simulations of AGN feedback interacting with the surrounding
turbulent and multi-phase ISM structure. Usually sub-grid models rely on a
number of assumptions regarding the coupling between the wind, jet, or
radiation with the gas. Such sub-grid models are not expected to generally
produce the same results as those found in resolved high-resolution AGN
feedback simulations. With an increasing number of simulations that manage to
resolve more of the the multi-phase structure of the ISM, it becomes
important to also ensure that AGN feedback models correctly bridge the gap
between large and small scales. It is therefore time for new simulations that
will use a physically validated approach in modelling AGN feedback to properly
understand the interaction with the multi-phase gas and how exactly AGN
feedback drives large-scale winds. Such simulations then can contribute to a
better understanding of how hydrodynamical sub-grid models can be improved in
light of the results. Because the properties of the ISM are so tightly linked to
star formation and stellar physics, the effort must go hand-in-hand with the
development of a stellar feedback model that regulates star formation as well
as the properties of the ISM such that it matches with observations.

The goal of the \textbf{\satin{}} (\textbf{S}imulating \textbf{A}GNs
\textbf{T}hrough \textbf{I}SM with \textbf{N}on-Equilibrium Effects) project is
to improve our understanding of the detailed interaction of the AGN with the 
turbulent multi-phase gas and how large-scale winds are driven. The entire galactic MW disc simulations have sufficient resolution to capture a distinct multi-phase ISM with the different gas phases interacting with each other. The simulations cover an order of magnitude in gas surface density, and naturally takes into account shear and radial motions. 

This is the first paper of the project where we present the initial building blocks of the simulations, in particular the star formation model via sink particles as well as the SNe feedback implementation. The \satin{} model incorporates a successful star formation and stellar feedback model that regulates star formation as well as the properties of the ISM. The SNe feedback model uses
a stellar evolution model and single star tracking to get individual SN delay times for the massive stars. Additionally, it adaptively adjusts to the local 
environment of each SN explosion by switching between the injection of thermal
energy and momentum depending on the surroundings. We then test in this paper the model in a full self-gravitating MW-like disc galaxy simulation, quantify the interaction of the SNe with the multi-phase and turbulent ISM, and compare the results with observations. Only once we established that the used stellar feedback model fulfils our requirements of regulating the properties of the gas within the galaxy as well as star formation we then go forward and analyse in detail the interaction of the AGN with the simulated galaxy. This will be subject of future papers. 

The paper is structured as follows: In Section~\ref{sec:SatinModel}, we
introduce the first building blocks of the \satin{} model and then explain in
Section~\ref{sec:setupDescription} the simulation setup of the isolated
self-gravitating turbulent MW-like disc galaxy that we simulate with and
without SNe feedback. In Section~\ref{sec:results}, we first give a qualitative
overview of the global evolution and morphology of the simulations.  We then
describe the star formation rates of the galaxies and compare them against
observations in Section~\ref{sec:SFR-KS}. In Section~\ref{sec:ISM} we then
investigate the multi-phase ISM structure and the mass and volume-filling
factors of the gas within the galaxy.  Furthermore, we look at the turbulent 
structure of the ISM in
Section~\ref{sec:Turbulence} and quantify the galactic outflows in
Section~\ref{sec:gasDynamicsFlows}.  In Section~\ref{sec:Discussion} we discuss
the results of our simulations in the context of other studies as well as
possible caveats. Finally, the paper is summarised and concluded in
Section~\ref{sec:conclusion}. In the
Appendix~\ref{sec:PowerSpectrumCalculations} we describe in more detail the
calculations of the power spectra presented in this paper.  
\section{SATIN Model}
\label{sec:SatinModel}
In this Section we introduce the first incarnation of our \satin{} model that we use
to study the ISM within a simulated MW-like galaxy. The model is a galaxy wide 
implementation of a successful ISM feedback model tested in small box simulations \citep[see]{Walch+2015} and adapted to the adaptive mesh refinement (AMR) code \Ramses{}
\citep{Teyssier2002}. 
We will begin with a summary of the numerical methods used to model gravity and hydrodynamics
(Section~\ref{sec:NumericalMethod}). Then we describe the physical models
such as non-equilibrium cooling (Section~\ref{sec:Cooling}), the formation of
massive stars (Section~\ref{sec:SinkFormation} and
\ref{sec:MassiveStarFormation}), a stellar stellar evolution model to track the age and death of the massive stars (Section~\ref{sec:StellarEvolution}),
and the implementation of SN feedback with individual delay times (Section~\ref{sec:SNeFeedback}). 

\subsection{Numerical Methods}
\label{sec:NumericalMethod}
For the simulations we use the AMR hydrodynamics code \Ramses{}. The code solves the poisson equations for collisionless particles (DM, stars, sinks) 
coupled to the hydrodynamics of an inviscid fluid via an AMR finite volume method. The motions of the collisionless particles are
evolved through the gravitational force with an adaptive particle mesh solver
using a cloud-in-cell interpolation, taking into account the mass contribution
from the gas. The gas is modelled with a second-order unsplit Godunov scheme.
We use the HLLC Riemann solver \citep{Toro+1994} with MinMod total variation
diminishing scheme to reconstruct the interpolated variables from their
cell-centred values. To relate the pressure and internal energy we use an 
adiabatic index of $\gamma = 5/3$. 
\subsection{Non-equilibrium Cooling}
\label{sec:Cooling}
The gas temperature and the non-equilibrium ionisation states of hydrogen and
helium are tracked using the \textsc{ramses-rt} radiative hydrodynamics (RHD) extension to
\textsc{ramses}. In the simulations presented we only use the chemistry module of \textsc{ramses-rt} and no photons
are created and transported. The details of this method, in particular the ionisation
chemistry, are given in \cite{Rosdahl+2013} but we summarise the processes
here.  We assume a hydrogen mass fraction of $0.76$ and a helium mass fraction of
$0.24$.  The cooling and heating rates include collisional ionisation and
excitation \citep{Cen1992} and formation of hydrogen and helium 
by recombination
\citep{Hui+Gnedin1997}.  Also included are
Bremsstrahlung cooling \citep{Osterbrock+Ferland2006}, dielectric recombination
\citep{Black1981} and Compton electron scattering off cosmic microwave background
photons \citep{Haiman+1996}.  We include hydrogen and helium photo-ionisation
and heating from a UV background at redshift zero \citep{FG+2009}.
Additionally we enforce an exponential damping of the UV radiation above the
self-shielding density of $n_H = 10^{-2}$~cm$^{-3}$. For the gas with
temperatures above $10^4$~K, the contribution to cooling from metals is added
using Cloudy \citep{CLOUDY1998} generated tables. It assumes photoionisation
equilibrium with a reshift zero UV background \citep{Haardt+Madau1996}.  Below
a gas temperature of $10^4$~K we use fine structure cooling rates from
\citet{Rosen+Bregman1995} which allow the gas to radiatively cool down to
10~K. The thermochemistry of molecules is not included in the public
version of \textsc{ramses-rt} that we use.    
\subsection{Sink Formation}
\label{sec:SinkFormation}
Once a molecular cloud is formed and accumulated enough mass it collapses and
forms stars and star clusters. We employ collisionless sink particles to model
the formation of internally unresolved star clusters in dense regions that
undergo gravitational collapse. To find the formation sites of the sink
particles we detect high density clumps by running a clump finder
\citep{Bleuler+Teyssier2014} at every global timestep. This method identifies all peaks and their
highest saddle points if the density of the gas cell is above a set threshold
density (i.e. $\rho_{sink} = 50$~H~cm$^{-3}$).  
When the peak-to-saddle ratio is greater than $1.5$
we recognise a clump as an individual entity, whereas otherwise we merge the
density peak with the neighbour peak with which it shares the highest saddle
point. We then investigate the gas surrounding the density peak for
gravitational collapse. We form a sink particle if a number of requirements are
met. First, we perform a virial theorem type analysis to ensure that the
gravitational field at a possible location for sink formation is strong enough
to overcome internal support of the gas within the clump. This avoids forming
sinks in gas which is only compressed by thermal pressure rather than gravity.
Secondly, we ensure that the requirement for a converging flow ($\nabla \cdot
\vec{v} < 0$) is met \citep[similar to][]{Federrath+2010}, ensuring that the
gas within the accretion volume contracts at the moment of formation.  Finally,
we do not allow the accretion radius to overlap with that of another existing
sink \citep[similar to][]{Federrath+2010}. If these conditions are met, we
form a sink particle.

Once a sink particle is formed it can, at each timestep, accrete gas from cells
which are within their accretion radius ($R_{acc} = 2 \times \Delta x$) and whose density is greater than the threshold density for formation. Additionally, we require the gas in a cell to fulfill the same conditions as for sink formation. Here $\Delta x$ is the distance from the sink to the cell centre of adjacent cells. Note, that we enforce the highest refinement within a sphere with radius $R_{res}=6\times\Delta x$ centred around each sink particle, ensuring that the accretion region has a well defined shape and that the SNe expands into a maximally refined region. 

If the requirements are met the accreted gas mass from a cell is 
\begin{equation}
\Delta m = \max \left(0.5(\rho - \rho_{sink}) (\Delta x)^3,0 \right) \, , 
\end{equation}  
where $\rho$ is the mean density within the accretion region. The code accounts for overlapping accretion regions by reducing the corresponding weight of the accretion mass considering the volume of the overlapping regions. All the formed
sink particles stay active throughout the whole simulation time. 

The formation and accretion of gas onto a sink particle at the chosen density will
reduce the local density and ensures that the Jeans length is resolved for
most of the gas cells within the simulations. This helps avoid articifial
collapse of the molecular cloud \citep{trueloveetal97}.  

The sink particles are evolved through the graviational force with a particle
mesh method already present in \Ramses{} for the DM particles (see
Section~\ref{sec:NumericalMethod}). Using this method is desirable due to the large
number of sink particles in our simulations. 
\subsection{Massive Star Formation}
\label{sec:MassiveStarFormation}
Typically sink particles are introduced to represent gravitationally collapsed
objects whose physical size is below the grid scale by orders of magnitude. In
our simulations the formed sink particles are considered to be tracing star
clusters and we use them to trace the most physical location for star formation
within the star cluster. We track the mass of every sink particle, which we
then define as the cluster mass.  We are only interested in the evolution of
massive stars (i.e, stars above 9~\Msolar) as those stars produce significant
amounts of ionising radiation, have powerful stellar winds, as well as
explode as SNe at the end of their lifetime. To follow the evolution state of
such massive stars that form within the star cluster we implement a sub-grid
model.  

First, we assume that all gas accreted onto the sink particles is converted
into stars. Each sink particle (star cluster) tracks the amount of mass
accreted onto it and every time this exceeds 120~\Msolar for an individual sink
particle we create a virtual object representing a massive star of mass $M$ and
attach it to the corresponding sink particle. We then decrement the accretion mass
by 120~\Msun~and repeat the process. By subtracting 120~\Msun~rather 
than the mass of the stellar object we account for the stars below
9~\Msun~in the mass distribution of the sink. We do not follow the low-mass
stars individually and thus assume that they do not emit any winds, radiation,
or explode in a SNe. The star itself is a virtual object that moves with the
sink and the number of massive stars associated with each sink differs. 
The object itself tracks the initial mass and age of the star, that is
then used by the stellar evolution and feedback model described in the next
Section. The initial mass of every new-born star is drawn from a  
Salpeter IMF \citep{Salpeter1955} within a mass range of 9-120~\Msolar. The same model specifics have been used in \cite{Gatto2017} and \cite{Peters+2017} for stratified ISM simulations. A
similar implementation can be found in \cite{Iffrig+Hennebelle2015} and
\cite{Geen+2018}. 
\subsection{Stellar Evolution}
\label{sec:StellarEvolution}
We follow the stellar evolution of the massive stars by tracking the age of the
star associated to a sink particle (star cluster) in order to get realistic SNe
delay times for each individual star within the star cluster. We use the
stellar evolution tracks \citep{Ekstroem2012} from the zero-age main sequence
(ZAMS) with a rotation rate $v{ini}/v_{crit}=0.4$. The evolution is computed
until the Wolf-Rayet/pre-SN phase. As already done in \cite{Gatto2017} we store
a grid of 112 tracks for stars in the mass range of 9-120~\Msolar, separated by
1~\Msolar. We interpolate linearly between tracks for each individual star.  We
make the simplified assumption that each formed massive star immediately starts
with the ZAMS evolution and thus do not account for a delay time due to star
formation or a proto-stellar phase (see~\citealp{Grudic+2022} as well as
\citealp{Verliat+2022} for simulations that includes protostellar jet feedback).
We assume that each massive star explodes as a Type \textsc{ii} SN once it has reached the end of its lifetime. The energy released by a
single SN event ($E_{SN} = 10^{51}$~erg) is injected into the surrounding of the
sink particle in the form of thermal energy or momentum input, depending on
whether the adiabatic phase of the SN remnant is resolved (see next
Section~\ref{sec:SNeFeedback} for detailed information). We also add the mass
of the SN progenitor to the injection region and decrement the total mass of
the sink by the same amount.  

In the \SATIN{} model we do not account for the unresolved stellar remnants or
runaway stars (see~\citealp{Andersson+2020} and \citealp{Steinwandel2022} 
for a detailed study of the effect of
runaway stars). Moreover, we did not include Type~\textsc{I}a SNe
explosions originating from an old stellar population.
\subsection{SN Energy and Momentum Input}
\label{sec:SNeFeedback}
The implemented SN model uses the individual delay times from the stellar evolution model and adaptively adjusts to the local environment of each SN explosion and switches between the injection of thermal energy and momentum depending on the surroundings. We release thermal energy
of $E_{SN}=10^{51}$~erg to the neighbouring cells of the star provided that the
adiabatic phase of the SN remnant is resolved. If, however, the density in the
injection region is high and the Sedov-Taylor phase is unresolved, we switch to
a momentum input scheme based on \cite{Blondin+1998}. 

For both models the thermal energy or momentum is distributed in a volume
weighted fashion within a sphere with radius $R_{inj}=2 \times \Delta x$. Note,
that the high resolution region around the star is larger (i.e $R_{res}=6
\times \Delta x$) and that the SNe thus expands into a maximally refined region
independent of the surrounding density. By doing so, we additionally avoid the
problem of dealing with coarse-fine boundaries between two refinement levels.  

We assume that each SN remnant ejects the mass of the progenitor star into its
surroundings and each cell in the injection region thus receives momentum
associated to the deposition of the ejecta from a star that moves with respect
to the mesh. In order to decide whether we inject thermal energy or momentum we first
calculate the mean local hydrogen number density $n_H$ in the region within
the injection radius and then calculate the radius of the bubble at the end of
the Sedov-Taylor phase as done in \cite{Blondin+1998}:
\begin{equation}
R_{ST} = 19.1 \, E_{51}^{5/17} \, n_H ^{-7/17} \, \, \mathrm{pc} \quad ,
\end{equation}
where $E_{51} = E_{SN} / (10^{51} \mathrm{erg} )$ is the number of SNe
exploding at the given time. If $R_{ST} < R_{inj}=2 \times \Delta x$ we inject
momentum rather than thermal energy. This ensures that, if we inject thermal
energy, the local cooling radius is sufficiently resolved by more than three
resolution elements along each Cartesian axis (see also \citealp{Kim+Ostriker2015} and \citealp{Martizzi+2015} for 
a discussion about this).

After the injection of thermal energy or momentum we update the timestep
within the simulation using the Courant-Friedrich-Levy stability condition
\citep{CFL1928}, where the timestep cannot be larger than
\begin{equation}
\Delta t = C_{CFL} \frac{\Delta x}{\max \left( \left| v \right| + c_s \right)} \quad . 
\end{equation}
Here $\Delta x$ is the cell width, $v$ the gas velocity, and  $c_{\mathrm{s}}$
is the sound speed. This is done in order to best capture the evolution of the
blast wave. We will now explain the two injection schemes in more detail.
\subsubsection{Thermal energy input}
If the SN is resolved we inject the total energy of $E_{SN} =
10^{51}$erg per SN to all the cells within the injection radius (i.e
$R_{inj}=2 \times \Delta x$). The thermal energy in each cell of the 
injection sphere is therefore updated as:
\begin{equation}
\epsilon_{th,cell} = E_{SN} / V_{fb} 
\end{equation} 
where $V_{fb}$ is the volume of all the cells within the injection region.

Along with the thermal energy we also inject the ejected gas mass, while
conserving momentum, into the cells within the same region. Here, we assume
that each SN remnant ejects the mass of its progenitor star into the
surroundings. And finally, we update the total energy of the gas cell taking
into account the changes to the thermal and kinetic energy. 

This can increase the temperature of a cell up to $\sim 10^8$~K in a
low-density medium where the injected mass is much higher than the mass of the
surrounding. The over-pressured gas then expands into the ambient,
inhomogeneous, ISM gas as a Sedov-Taylor blast wave.  In principle about
$\sim$~30\% of the total thermal energy can be deposited into kinetic energy
\citep{Chevalier1974}. However, as is already well known \citep{Katz1992,
NavarroWhite1993, Abadi+2003, Slyz+2005, Stinson+2006, Creasey+2011,
Hummels+Bryan2012, Kimm+Cen2014} the atomic and metal cooling processes in the
gas can rapidly radiate the internal energy away before the blast wave sweeps
up the ambient medium in dense environments or in simulations where the cooling
radius is under-resolved. This is why we switch to a momentum-input scheme in
situations where the cooling radius is unresolved.  
\subsubsection{Momentum input}
In the case of an unresolved Sedov-Taylor phase of the SN remnant we follow a
momentum-input scheme. The input momentum is at solar metallicity calculated as
\citep{Blondin+1998, Thornton+1998, Kim+Ostriker2015, Geen+2015}
\begin{equation}
p_{ST} = 2.6 \times 10^5 E_{51}^{16/17} n_H^{-2/17} \Msun \mathrm{km} \mathrm{s}^{-1} \quad . 
\end{equation}
We inject the momentum taking into account that each progenitor star particle
moves with respect to the mesh, and we assume that each SN remnant is
spherically symmetric in the frame of reference of the progenitor star. We
further assume that the centre of each SN is the position of the parent sink
particle associated with the exploding progenitor star. The injected momentum 
is calculated using a velocity of
\begin{equation}
v_{inj} = \frac{p_{ST}}{M_{inj}} \hat{r} 
\end{equation}
where $M_{ej}$ is the ejected SNe mass (i.e mass of progenitor star), and
$\hat{r}$ is a unit vector that points from the sink particle position towards
the centre of the cell into which we inject the velocity. The velocity $v_{inj}$
thus points radially outwards. Along with the momentum from the SNe we also add
the ejected gas mass into the cells, again while conserving momentum. And as 
in the thermal-energy implementation we also change the total energy of the 
system taking into account the changes to the thermal and kinetic energy. 
\section{Simulation Setup}
\label{sec:setupDescription}
We explore our model in high-resolution simulations of an
isolated self-gravitating turbulent MW-like disc galaxy to naturally account for shear and radial motions. To test the effect
of the SNe in regulating star formation and on the simulated
multi-phase ISM we run the galaxy with (\SNe) and without SNe feedback (\nSNe).
We collectively refer to both simulations as the \satinI{} simulations. The simulated galaxy consists of a disc of gas and stars, a stellar bulge, and a dark matter (DM) halo. After two
initialisation stages described below we use a sink formation algorithm to model the formation,
evolution and eventual SNe explosion of massive stars within the star cluster
using stellar tracks.
\subsection{Initial Conditions}
We study an isolated self-gravitating turbulent MW-like galaxy made of a
disc of gas and stars, a stellar bulge, and a dark matter (DM) halo. The
initial conditions are set-up with the initial condition code DICE
\citep{DICE2016}. The DM halo initially follows a NFW \citep{NFW96} density
profile with a concentration parameter of $c=22$. The virial velocity of the DM
particles is set to be $v_{200}=134$~km/s, which corresponds to a virial radius
of $R_{200} \approx 190$~kpc and a virial mass of $M_{200}=79 \times
10^{10}$~\Msun .  With a cutoff radius of 12~kpc the total mass of the DM halo
is $3.6 \times 10^{11}$~\Msun.  We use $10^{6}$~DM particles to sample the
halo.

The total stellar mass of the galaxy is $4.6 \times 10^{10}$~\Msun and the
total gas mass of the galaxy is $0.59 \times 10^{10}$~\Msun, resulting in an
initial gas fraction within the galaxy of $f_{gas} = 0.12$.  The stellar and
gas disc follow an exponential and sech-z profile with a scale length of 6~kpc
and scale height of 0.4~kpc.  The stellar bulge with a total mass of $1.2
\times 10^{10}$~\Msun follows a Hernquist profile \citep{Hernquist1990} of
scale radius 2~kpc. We use $1.5 \times 10^6$ and $2 \times 10^5$ star particles
to sample the disc and bulge, respectively. The stellar mass resolution is $2.7
\times 10^4$~\Msun. 

The galaxy is initialised with a uniform solar metallicity ($Z=0.014$; \citealp[][]{Lodders+2009}) and no
metals are placed outside the disc. The circumgalactic medium (CGM) initially
consists of a homogeneous hot and diffuse gas with a constant hydrogen number
density of $n_{CGM}=10^{-6}$~H~cm$^{-3}$, temperature $T=10^6$~K, and zero
metallicity.  The advection of metals is tracked as a passive scalar on the AMR
grid. 

Before we turn on all the physics discussed in detail above
(Section~\ref{sec:SatinModel}) we go through two initial phases; a relaxation
phase and turbulence injection phase. In the relaxation phase we allow the
galaxy to relax to an equilibrium configuration (with a reasonable disc
thickness) for 50~Myr. We perform this first phase without gas cooling, sink
formation, and feedback. It allows removing signatures from the imperfect
equilibrium of the initial conditions.  After this first relaxation phase we
turn on gravity and gas cooling and randomly explode stars in the galaxy for
another 10~Myr to inject additional turbulence and to prevent the galaxy from
collapsing into a thin disc (similar approach has been used in \citealp{Hu+2016}).  
The stars explode at a random location within the disc and at a rate expected from 
the surface density of the galaxy and assuming a Kennicutt-Schmidt (KS) relation
\citep{Kennicutt1998}.  Only after these two stages we turn on sink formation
to follow the formation and evolution of the star clusters formed within our
galaxy self-consistently.  The stars set in the initial condition of the galaxy
contribute only to the dynamical evolution and gravitational potential of the
rotating disc, but they never explode as SNe. 

\subsection{Adaptive Refinement}
The box size is 650~kpc with the coarsest level of 9 and a
maximum level of 16 corresponding to a minimum cell size of 9~pc for most of
the galaxy. For refinement we employ a quasi-Lagrangian scheme where a cell is
refined if the gas within a cell is larger than $5 \times 10^3$~\Msun or if the
cell has eight or more DM and/or star particles within the cell. Additionally,
we also refine at each level if the cell size exceeds 80\% of the local
Jeans length, until the region reaches the maximum resolution of 9~pc
(corresponding to level 16).  Note that at maximum resolution, the Jeans length
can become under-resolved for a few cells which can lead to artificial collapse
of self-gravitating gas~\citep{trueloveetal97}.  We thus set the sink formation
density threshold to a value (i.e $\rho_{sink}$=50~H~cm$^{-3}$) at which the
Jeans length becomes resolved by less than 4 cells in gas with temperatures
below 200~K at the highest refinement level (see also
Section~\ref{sec:SinkFormation}).  Forming stars from this high density gas
will reduce the local density and ensures that the Jeans length is resolved for
most of the gas cells.  Note that for a majority of the cells refinement is
triggered due to the gas mass scale rather than due to the Jeans length
criteria. This causes the galaxy to have much more refined cells than if only
the Jeans criterion is used. In addition, we also enforce the highest
refinement within a sphere with radius $R_{res}=6\times\Delta x$ centred around
each sink particle, ensuring that the SNe expands into a maximally refined.
We found that these two additions are important for
the SNe to be properly resolved as it ensures that the gas around the star is
also in a highly resolved region.
\begin{figure*}
\plotoneF{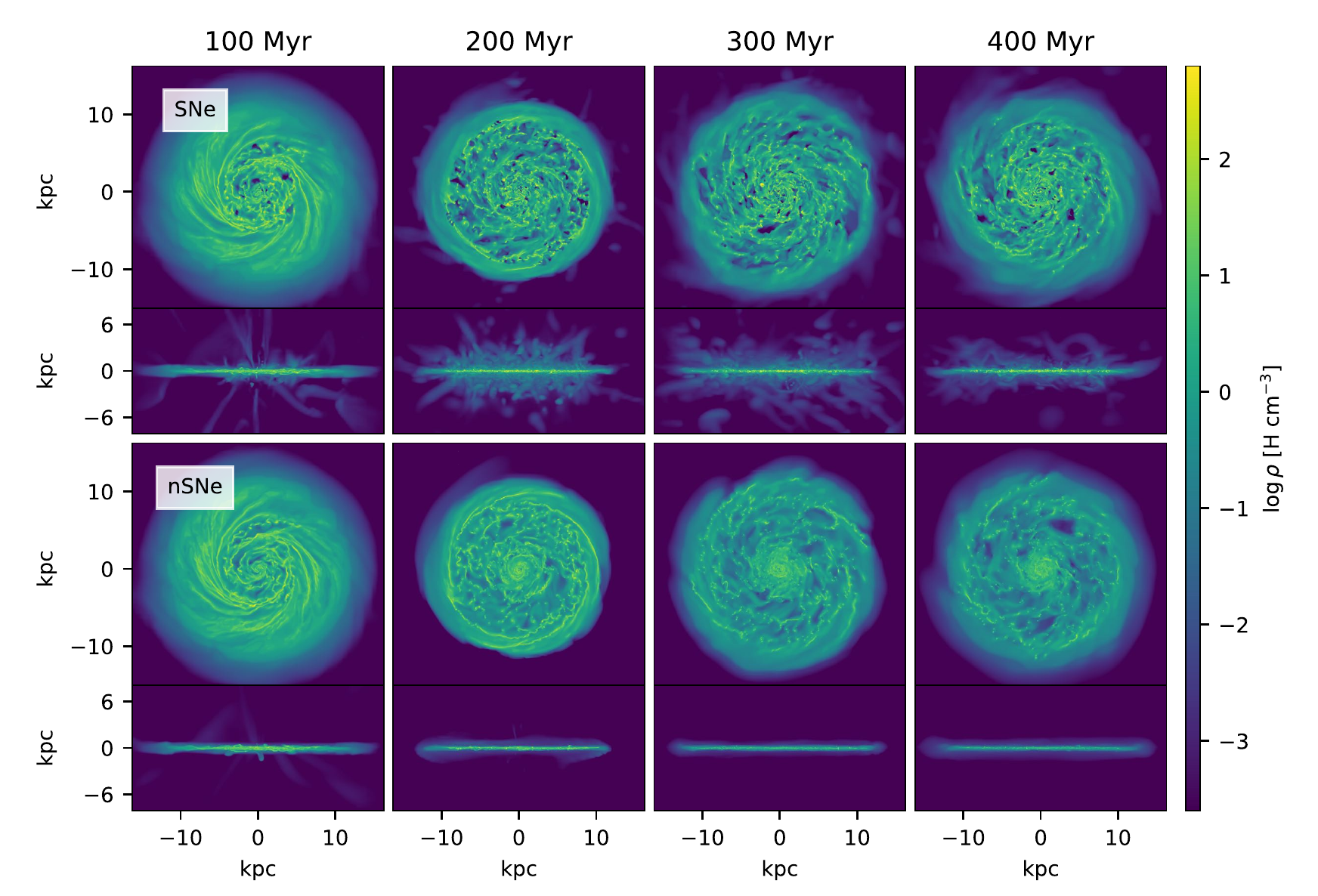}
\caption{Gas density maps (mass-weighted) of the \nSNe{} (top) and \SNe{}
(bottom) simulation for different times (t=100, 200, 300, 400~Gyr) from left to
right. Each panel shows both face-on (32$\times$32~kpc, upper part) and edge on
(32$\times$16~kpc, lower part) views. The pixel size is taken to be $dx=10$~pc.
The appearance of the ISM is smoother and no cavities of lower densities are
visible in the \nSNe{} simulation due to the lack of SNe. SNe creates a
multi-phase ISM structure within the whole disc and pushes the gas a few kpc
above and below the disc plane increasing the disc thickness.} 
\label{fig:GasDensityEvolution}
\end{figure*} 

%
%
\section{Results}\label{sec:results} 
%
\begin{figure*}
\plottwoH{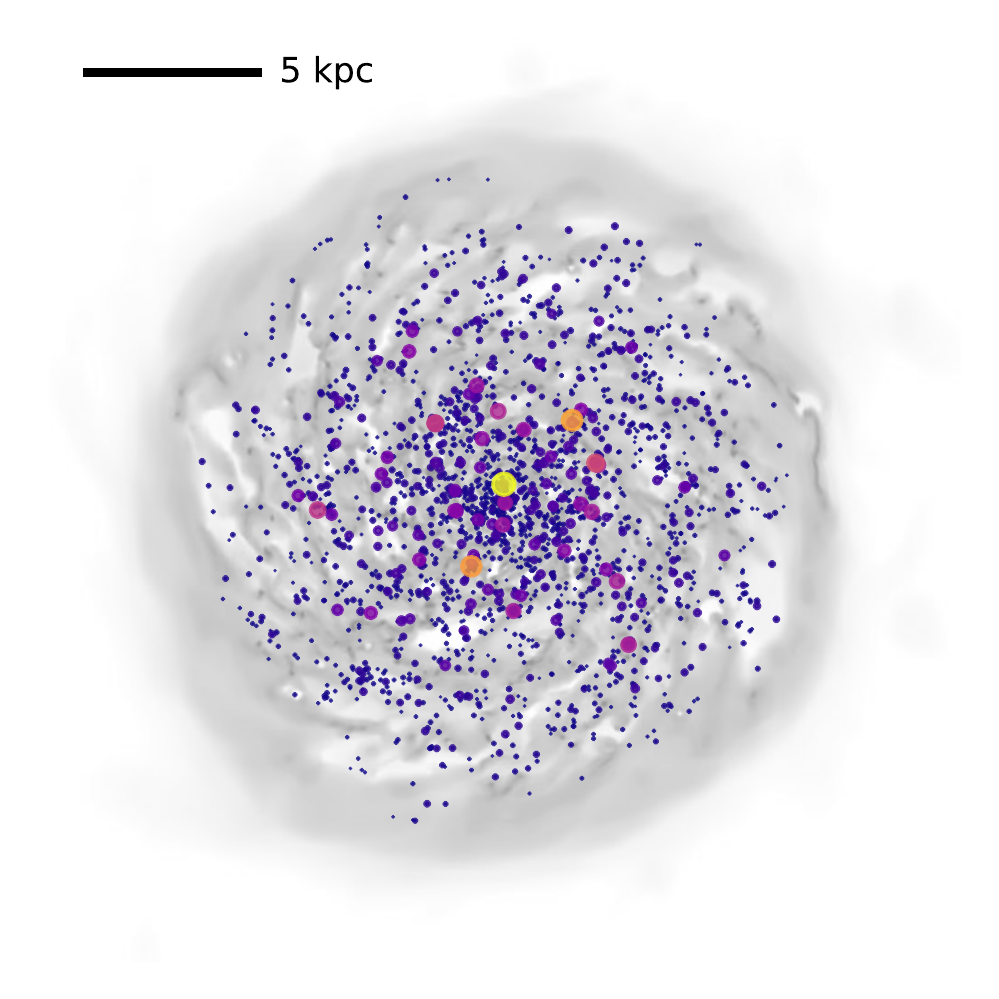}{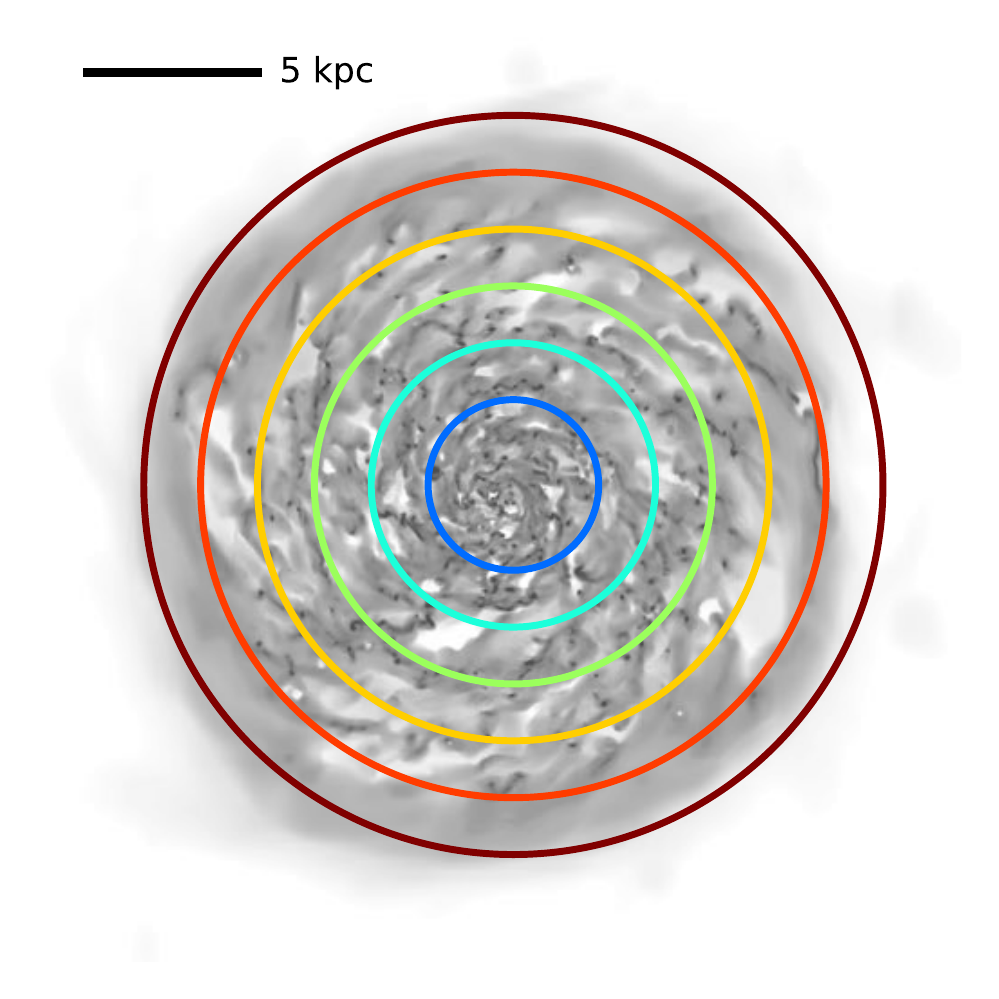}
\caption{\textit{Left:} Sink distribution within the galaxy at t=350~Myr for the 
\SNe{} simulation plotted over the corresponding gas density distribution. 
The size of each point is scaled
with the mass of the star cluster where more massive stellar clusters appear
larger.  The color of the points changes from blue (low mass star clusters) to
yellow (high mass star clusters). We chose the color and point size to best
highlight structures within the galaxy. For instance, a spiral structure
within the star cluster distribution starts to emerge in the outer regions of
the galaxy. Within the centre the star cluster distribution appear more
chaotic.  \textit{Right:} Gas distribution (mass-weighted) for the \SNe{} simulation 
at the same time t=350~Myr.  Overplotted are concentric rings ($r=3,5,7,9,11,13$) in the same
colors as used in Figure~\ref{fig:surfSFRrings}.}
\label{fig:SinkRingMaps}
\end{figure*}
\begin{figure*}
\plottwoH{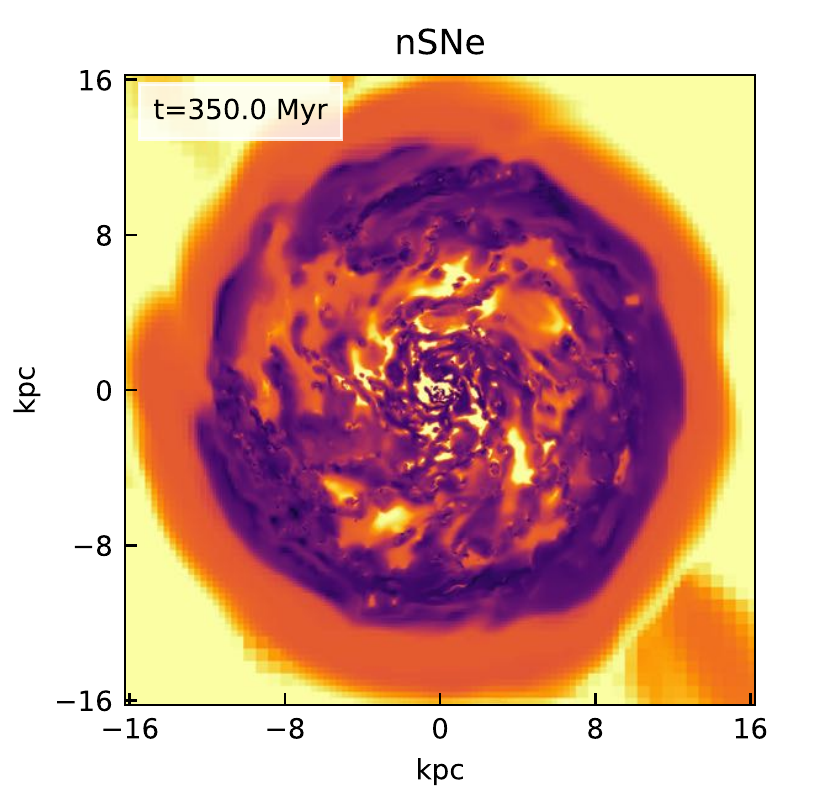}{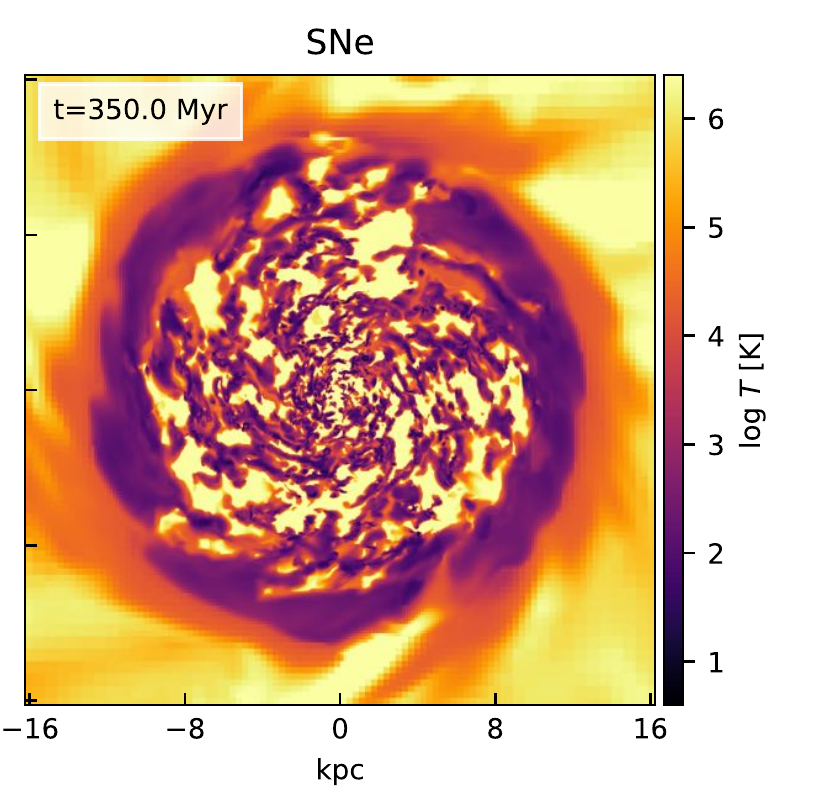}
\caption{Face-on density-weighted gas temperature slice of the \nSNe{} (left)
and \SNe{} (right) at t=350~Myr. The map scale and pixel size is as in
Figure~\ref{fig:GasDensityEvolution}.  The hot gas cavities in the \SNe{}
simulation is filled with shock-heated gas around $10^7$~K. A distinct three-phase
ISM is visible in the \SNe{} simulation. The dynamical evolution of the \nSNe{} creates
holes that is filled with hot gas from the surrounding. More volume is filled by the
warm than hot gas for the \nSNe{} simulation.} 
\label{fig:GasTemperatureMapComparison}
\end{figure*}
\begin{figure*}
\plottwoH{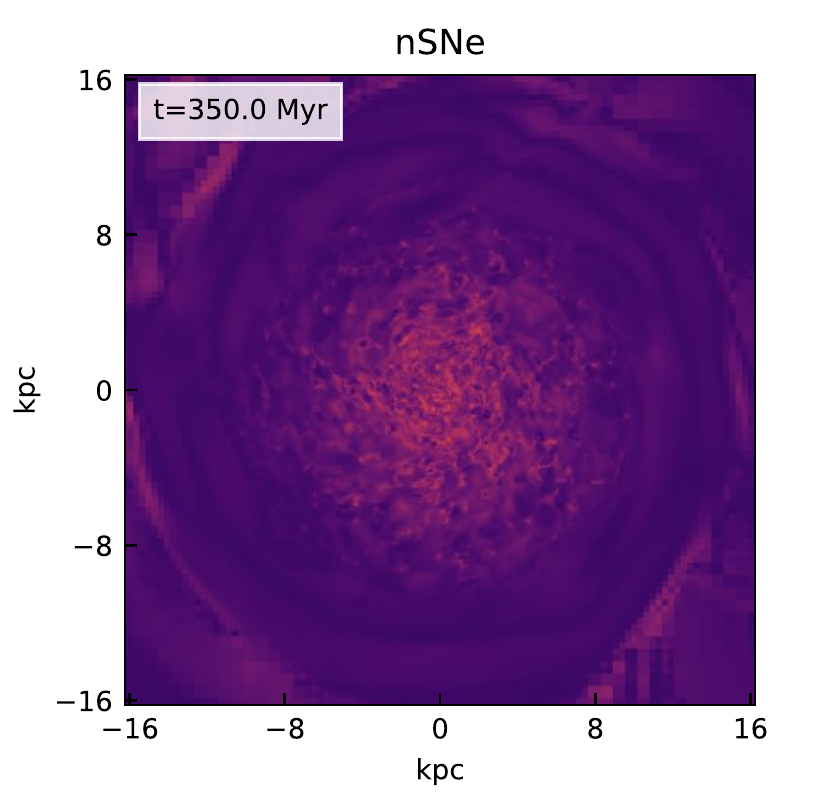}{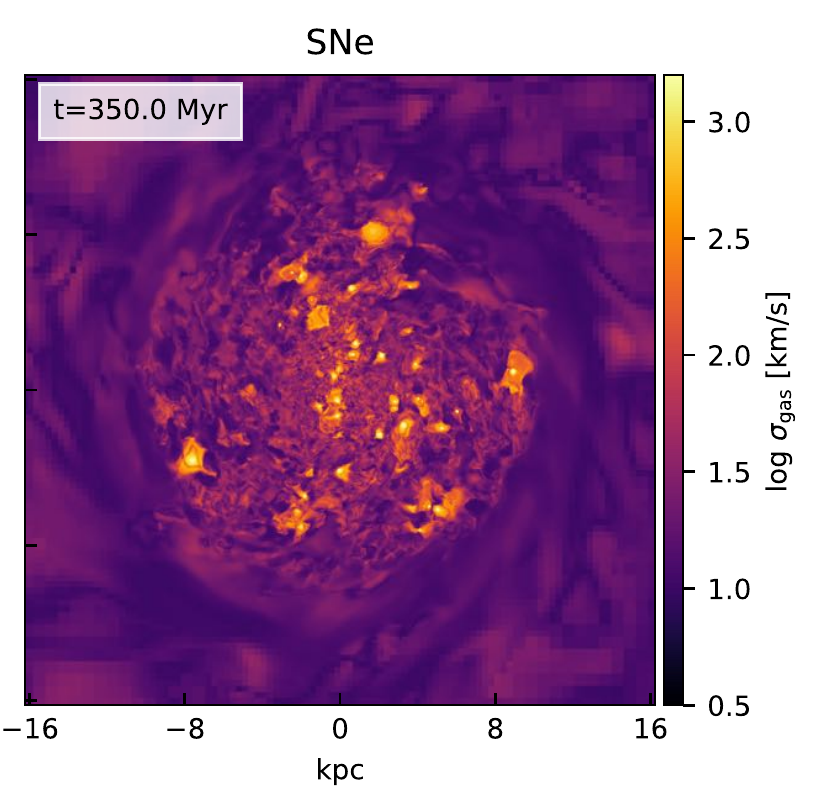}
\caption{Line-of-sight gas velocity dispersion maps (in km/s) of the \nSNe{} and
\SNe{} simulation at $t=350$~Myr. The map scale is as in
Figure~\ref{fig:GasDensityEvolution}. For each plotted pixel we calculated the
velocity dispersion within a beam of $\pm 100$~pc measured from the midplane.
The width of the pixels are taken to be $dx=10$~pc. Recent SNe clearly increase
the velocity dispersion within the low density gas of the galaxy indicating
that the multi-phase structure within the disc is not only driven by gravity and
shear but also stellar feedback events.}
\label{fig:VelocityDispersionMap}
\end{figure*}
In this Section we turn to the two simulated galaxies run with (\SNe) and
without SNe feedback (\nSNe) as the only difference between the two
simulations. We explore the ability of the \satin{} model to regulate star
formation and look whether or not the galaxy manages to reach a self-regulated
steady state (Section~\ref{sec:SFR-KS} and \ref{sec:ISM}).  To understand
whether the \satin{} model reasonably regulates the formation of stars we will
also compare the simulations with observational data. Further, we will
investigate the multi-phase ISM structure of the two simulations and examine
the impact of the SNe on the structure (Section~\ref{sec:ISM}) and energetics
(Section~\ref{sec:Turbulence}) of the ISM. Finally, we discuss the gas dynamics
and flows (Section~\ref{sec:gasDynamicsFlows}). 
\subsection{Qualitative Differences}
The evolution of the two simulated galaxies, \SNe{} and \nSNe{}, are visually
different from each other. Figure~\ref{fig:GasDensityEvolution} shows the
evolution of the gas density maps for the two different simulations.  The
initial 22~Myr of evolution are, after the two initialisation phases, identical
for the two simulations. The simulations begin to diverge only once the first
SN explodes and low density pockets appear due to the effect of the expanding
SNe bubbles. 

Stars form predominantly along the spiral pattern in the gaseous disc, where
the density is the highest. The first SNe explode in these high-density
regions, carving out cavities of low-density gas, and destroying the gaseous
spiral pattern within the galaxy (see evolution of the \SNe{} run in the lower
panel of the Figure).  After the first 200~Myr the gas distribution of the
\SNe{} run becomes increasingly structured again. This is because at that stage
the SFR is decreasing and thus there are less SNe disturbing the gas
distribution which allows the gas to settle more. Dense and cold gas regions
within which star clusters form are surrounded by hot, lower density gas and a
multi-phase ISM develops. Gas bubbles are pushed out by the SNe from the
star-forming disc into the halo. As we will see, part of the gas falls back
onto the disc while some gas escapes the galaxy. 
 
Comparing the two runs in Figure~\ref{fig:GasDensityEvolution}, we observe that
the gas in the \SNe{} simulation reaches both higher and lower density
extremes.  Due to the lack of SNe the appearance of the ISM in the \nSNe{}
simulation is smoother and no large cavities are visible. Continued gas
accretion onto the stellar clusters (sinks) removes gas from within the disc
and leads to star formation within the cluster. This alone already helps to
form structures within the ISM, as seen in the \nSNe{} simulation.  The central
region of the \nSNe{} simulation accumulates more gas over time, whereas SNe
explosions disrupt the central region of the galaxy and create a multi-phase
structure in the centre.   

The edge-on view show that in the \SNe{} simulation gas flows out of the galaxy
and the SN feedback increases the disc thickness compared to the \nSNe{}
simulation where the galaxy develops into a razor-thin disc. The large-scale
outflows generated by the SN explosions push the gas to a few kpc above and
below the disc plane. The outflows correlate with star formation and start in
the central region of the galaxy where the first stars are born.  At later
times (200~Myr) when star formation also reaches the edge of the galaxy,
outflows escape from the full disc plane (see
Section~\ref{sec:gasDynamicsFlows} for a quantitative discussion on the gas
flows). 

Figure~\ref{fig:SinkRingMaps} shows on the left side the sink distribution
within the galaxy at $t=350$~Myr for the \SNe{} simulation (overplotted onto the density distribution).
The point size is weighted by the mass of the star cluster. The color of the
points changes from blue (low mass star clusters) to yellow (high mass star
clusters). The size and color distribution shows that the high mass clusters
are mostly around the centre of the galaxy where the first stars form. A spiral
structure in the star cluster distribution starts to emerge in the outskirts of
the galaxy. On the right side the gas distribution is shown for the same
snapshot and same size. Overplotted are circles ($r=3,5,7,9,11,13$) in the same
colors used in Figure~\ref{fig:surfSFRrings}.  We will use the same ring radii
for different analyses later on. 

The face-on temperature slices in Figure~\ref{fig:GasTemperatureMapComparison}
at $t=350$~Myr show a complementary picture to the density maps.  The densest
regions contain the lowest temperatures for both the \SNe{} and \nSNe{}
simulation, whereas the highest temperatures can be seen in the low density
regions within the galaxy. A distinct three-phase ISM with cold ($T<300$~K), warm
($T=300-2 \times 10^5$~K), and hot ($T>2\times 10^5$~K) gas is visible in the \SNe{}
simulation for most of the galaxy. The gas structure
outside $\sim$~9~pc is at $t=350$~Myr still very smooth and cold. In this outer
region the gas did not yet form a multi-phase structure. The hot gas in the
\nSNe{} simulation originates from the hot surrounding of the galaxy that flows
into the low density holes created by the ever accreting sink particles. In the
\SNe{} simulation, the gas cavities carved by the accreting sinks as well as
SNe feedback is filled with shock-heated gas around $10^7$~K.  It is still
possible that part of this hot gas also originates from the hot surrounding
medium around the galaxy. Given that the ratio of outflow versus inflow rates
(OFR/IFR) discussed in Section~\ref{sec:gasDynamicsFlows} is well above one for
the majority of the simulation, we assume that this is, however, a small
amount. 

A large portion of the low density hot gas is also pushed outside the
disc by the SN explosions.  The temperature slice for the \SNe{} simulation
makes it apparent that the hot gas fills most of the galaxy volume, followed by
the warm and then cold gas. For the \nSNe{} simulation more volume is filled by
the warm than hot gas (see Section~\ref{sec:ISM} for a more quantitative
discussion of the volume filling fractions). 

Figure~\ref{fig:VelocityDispersionMap} shows the vertical gas velocity
dispersion for the \nSNe{} and \SNe{} simulation at a fixed time, $t=350$~Myr.
We measure the line-of-sight velocity dispersion map of the gas by projecting
the galaxy face-on and include for each pixel of the image cells within a beam
of $\pm 100$~pc measured from the mid-plane of the disc. The width of the 
pixel and beam is taken to be $dx=10$~pc. For each beam the line-of-sight
gas velocity dispersion is calculated as 
\begin{equation}
\sigma_{gas} = \sqrt{\frac{\sum _ i (v_{z,i} - \overline{v}_z)^2}{N}} 
\end{equation}
where $v_{z,i}$ is the line-of-sight velocity in pixel $i$, $\overline{v}_z$ is
the mean velocity within the beam, and $N$ is the number of pixels in each
beam. 

The velocity dispersion within the galaxy in the \SNe{} simulation is clearly
higher than for the \nSNe{} simulation. The very high velocity dispersion regions
correlate with the low density and hot temperature cavities carved by recent
SN events, but the velocity dispersion also seems to be higher in regions of
higher densities and lower temperatures.  

In conclusion, we have seen that with SN feedback the ISM becomes multi-phase
and cavities are formed from the explosions that are low density and hot. The
velocity dispersion is not only high within the cavities but also in warm gas.
This indicates that the multi-phase structure within the galaxy is not only
driven by gravity and shear but also stellar feedback events. The ISM
properties are further quantified in Section~\ref{sec:ISM}. The analysis of the
kinetic energy power spectrum in Section~\ref{sec:Turbulence} will additionally
show that SNe feedback is important for the energetics of the ISM. Especially
for the warm phase that is only turbulent when SNe are included.
\subsection{Star formation rates and the Kennicutt-Schimdt relation}
\label{sec:SFR-KS}
\begin{figure}
\plotone{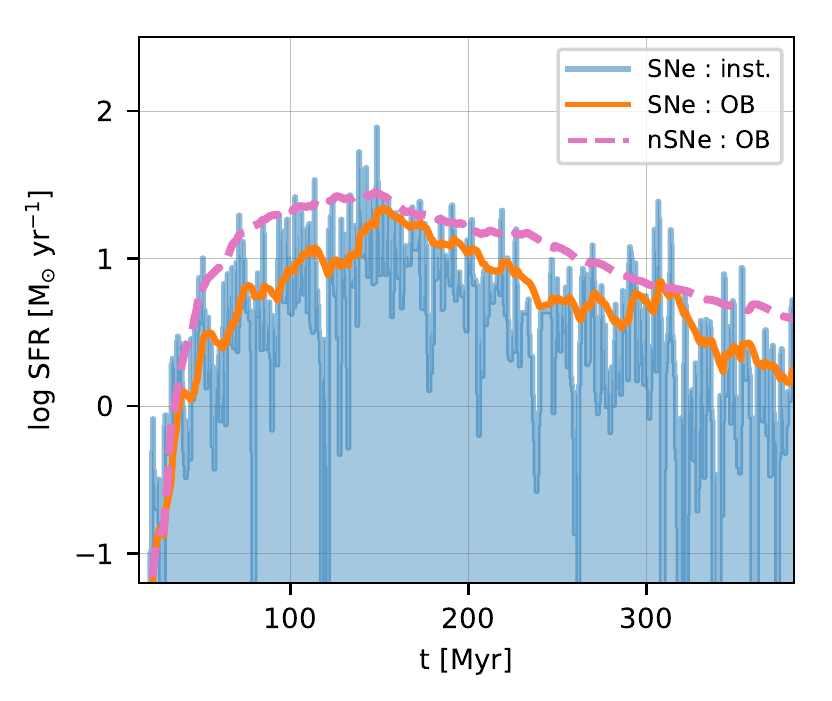}
\caption{Global SFR history for the MW-like galaxy. The light-blue bins
represent the instantaneous star formation rate for the \SNe{} simulation computed
as the sum of the individual instantaneous star formation rates of each
individual stellar cluster at a given time $t$ (see Eq.~\ref{eq:SFR_inst}). The
orange line shows the observed SFR values of the \SNe{} simulation derived from the
O- and B type star formation lifetimes (see Eq.~\ref{eq:SFR_OB}). The dashed
pink line shows the observed SFR values for the \nSNe{} simulation. The SFR of the
\nSNe{} simulation is smoother and higher by a factor of of two (as a mean value for the
length of the simulation) than the SFR of the \SNe{} simulation. SFR in both simulations
decreases with time after 100~Myr but the SFR is much more stochastic for the \SNe{}
simulation.}
\label{fig:totalSFRtime}
\end{figure} 
\begin{figure*}
\plotoneF{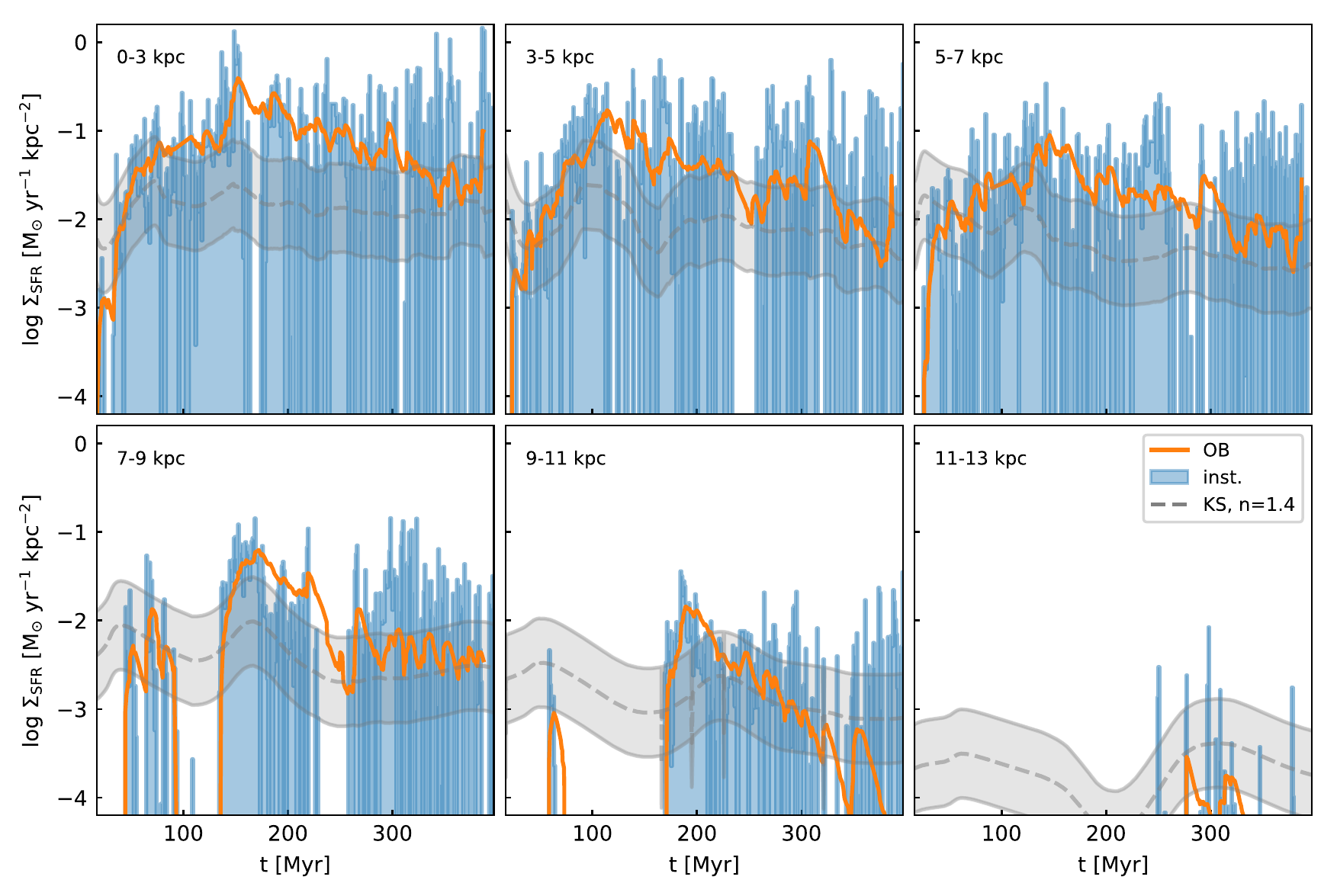}
\caption{Evolution of \sigmaSFR within different concentric cylindrical rings
(radius range written in figure panel). Blue is for
\sigmaSFR$_{\mathrm{inst}}$, and the orange line for
\sigmaSFR$_{\mathrm{OB}}$ for the \SNe simulation, respectively (as in
Figure~\ref{fig:totalSFRtime}). The grey dashed line represents the expected
\sigmaSFR from the Kennicutt-Schmidt relation for the calculated surface
density within the cylindrical ring at a given time. The grey band indicates a
factor of two uncertainty. For a visual impression of the different radii
within the gas density distribution at t=350~Myr see
Figure~\ref{fig:SinkRingMaps}.  Supernova feedback alone regulates SFR well in
lower surface density environments (i.e radii at and above solar
neighbourhood around 7-9 kpc). 
For higher surface densities SNe alone is, at early times, not
sufficiently limiting SFR. Early feedback from radiation, cosmic rays, and
stellar winds likely would help.}
\label{fig:surfSFRrings}
\end{figure*} 
The regulation of star formation is an essential aspect of any galactic star
formation and feedback model and a probe of the efficiency of the feedback
process. Without stellar feedback most of the gas within the galaxy would just
be consumed and form stars. Star formation would then occur at a much
higher rate and on shorter time-scales than observed. In this Section, we will
investigate the ability of the \SATIN{} model to regulate star formation.  We
look at different regions within the disc, probe different gas
surface densities, and explore how the SN explosions manage to regulate star
formation in these regions. We will discuss which regions of the galaxy reach
a self-regulated steady state due to SNe feedback. To call a region within a
galaxy self-regulated we specifically require the surface density of the star
formation rate (\sigmaSFR), the projected density of the gas (\sigmaGas), as
well as the volume filling fractions (VFFs) and mass filling fractions (MFFs) to remain
constant over a certain time period (for a detailed discussion on the last requirements
see Section~\ref{sec:ISM}). We then observe how well the measured \sigmaSFR{}
and \sigmaGas{} match the KS relation and compare to observational data.  

Figure~\ref{fig:totalSFRtime} shows the total star formation history of the two
simulated galaxies (\nSNe{} and \SNe). The blue bars show the instantaneous
star formation rate for the \SNe{} simulation computed as the sum of the star
formation rates of each individual stellar cluster found within the galaxy at a
given time $t$ 
\begin{equation}
\mathrm{SFR}_{inst} = \sum _{j=1} ^{N_{sink}} \dot{M}_{sink, j} \left( \Delta t \right) \quad [ \mathrm{M}_\odot \mathrm{yr}^{-1}], 
\label{eq:SFR_inst} 
\end{equation}
for $t - \frac{\Delta t}{2} < t < t + \frac{\Delta t}{2}$. We took a bin size of
$\Delta t = 1$~Myr to calculate the instantaneous SFR. This corresponds to
$\sim$~2700 time steps in the \SNe{} simulation (the typical time step is
370~yrs).   

The orange and dashed-pink line is an estimate of the observable SFR within the
\SNe{} and \nSNe{} simulations.  The plotted observable SFR takes into account
the respective lifetime, $t_{\mathrm{OB}}$, of each individual massive star
formed within the simulated stellar clusters. It therefore tries to mimic the
SFR an observer would measure when tracing the SFR with H$\alpha$ emission,
where the emission sensitively depends on the presence of OB and WR-stars,
something that we track directly in our simulations.  The observable SFR is
defined as 
\begin{equation}
\mathrm{SFR}_{\mathrm{OB}} = \sum _{i=1} ^{N_\star} \frac{120}{t_{\mathrm{OB,i}}} \quad [ \mathrm{M}_\odot \mathrm{yr}^{-1}],
\label{eq:SFR_OB} 
\end{equation} 
where $t_{form,i} < t < t_{form,i} + t_{\mathrm{OB,i}}$. Here $t_{form,i}$ is the
time of formation of massive star $i$, and $N_\star$ is the total number of
massive stars at time $t$ (taken from \citealp{Gatto+2015}). 

For the first 150~Myr the SFR of the \SNe{} simulation rises steadily due to
the fragmentation of the initial gaseous disc and the subsequent formation of
stars within the galaxy. At this stage the galaxy has consumed only about 5\%
of the total gas mass present at the beginning of the simulation.  Following
this `star-burst' phase, the SFR decreases for the rest of the simulation.
This is due to three important aspects: Firstly, the rapid formation of stars
at the beginning of the simulation quickly removes dense gas around the star
clusters.  Secondly, the SNe additionally increase the turbulence within the
gas in the galaxy.  Thirdly, the SNe manage to push out gas from the galaxy.
All of this leads to a decline in gas accretion onto the star clusters and
results in a reduction of the newly formed stars.  Towards the end of the simulation the \SNe{} simulation approaches 3~\Msun/yr, which is in agreement with observations of this galaxy type. 
However, given that its SFR is still
decreasing at the end of the simulation we note that the whole galaxy has
not (yet) reached a completely self-regulated steady state.   

The \nSNe{} (dashed pink) and \SNe{} (orange) simulations only start to differ
after the first SNe explode around 22~Myr, where the SFR rises more steeply for
the \nSNe{} until around 150~Myr. However, the \nSNe{} galaxy consumed at this
stage already 30\% of the original total gas mass within the galaxy, compared
to 5\% for the \SNe{} simulation. This shows that the SNe reduce the accretion of
gas onto the star cluster already in the star-burst phase and subsequently
lower the rate at which stars are formed early on.  After this first rise, the
SFR of the \nSNe{} declines steadily until the end of the simulation,
continuously consuming gas. The constant disruption of the gas distribution
around the star clusters manifests in much more SFR variation on short timescales compared to the
\nSNe{} simulation.  

Turning on SN feedback reduces the formation of stars by a factor of two as a
mean value over the duration of the simulation. After 400~Myr the \nSNe{}
simulation consumed more than 80\% of the initial gas mass within the galaxy
and turned it into stars. On the other hand, the \SNe{} simulation only consumed
60\% of the gas mass due to the presence of the SNe feedback (see also 
Figure~\ref{fig:SFRResolution} for the evolution of the total mass of newly formed
stars). SNe feedback
induces turbulence within the gas as well as pushes gas outside the galaxy,
both reducing gas accretion onto the star clusters and the formation of stars
within them. Part of the ejected gas is recycled and used to sustain star
formation at later times.

As the global SFR is still decreasing at the end of the simulation
we conclude that the galaxy has not (yet) reached a self-regulated state.  
We will now look at different regions within the galaxy and probe the
capability of SNe feedback to regulate SF in different gas surface densities
environments. For this we will analyse the Kennicutt-Schmidt (KS) relation
\citep{Kennicutt1998} that connects the projected density of the gas
(\sigmaGas) to the surface density of the star formation rate (\sigmaSFR). 

Figure~\ref{fig:surfSFRrings} shows the SFR surface density within different
rings in the galaxy as a function of time.  Similar to
Figure~\ref{fig:totalSFRtime} the blue bars show the instantaneous star
formation rate surface density and the orange line shows the observable SFR
surface density evolution defined in Eq.~\ref{eq:SFR_OB} within the indicated
rings. A visual guide of the ring boundaries is shown in
Figure~\ref{fig:SinkRingMaps}.  Overplotted in a grey dotted line is the
calculated \sigmaSFR assuming that it follows the universal KS relation,
\sigmaSFR~$\propto$~\sigmaGas$^{1.4}$, where \sigmaGas is the measured gas
surface density at the given time.  We calculated \sigmaGas taking into account
all gas between $z=\pm 250$~pc and within the respective ring boundaries
centred on the galaxy\footnote{We tested that by using a larger delta $z$ we do not change the results significantly as there is not much mass above these distances}. The grey band shows an observational uncertainty of
1~dex around \sigmaSFR$(t)$.  

\begin{figure*}
\plotoneF{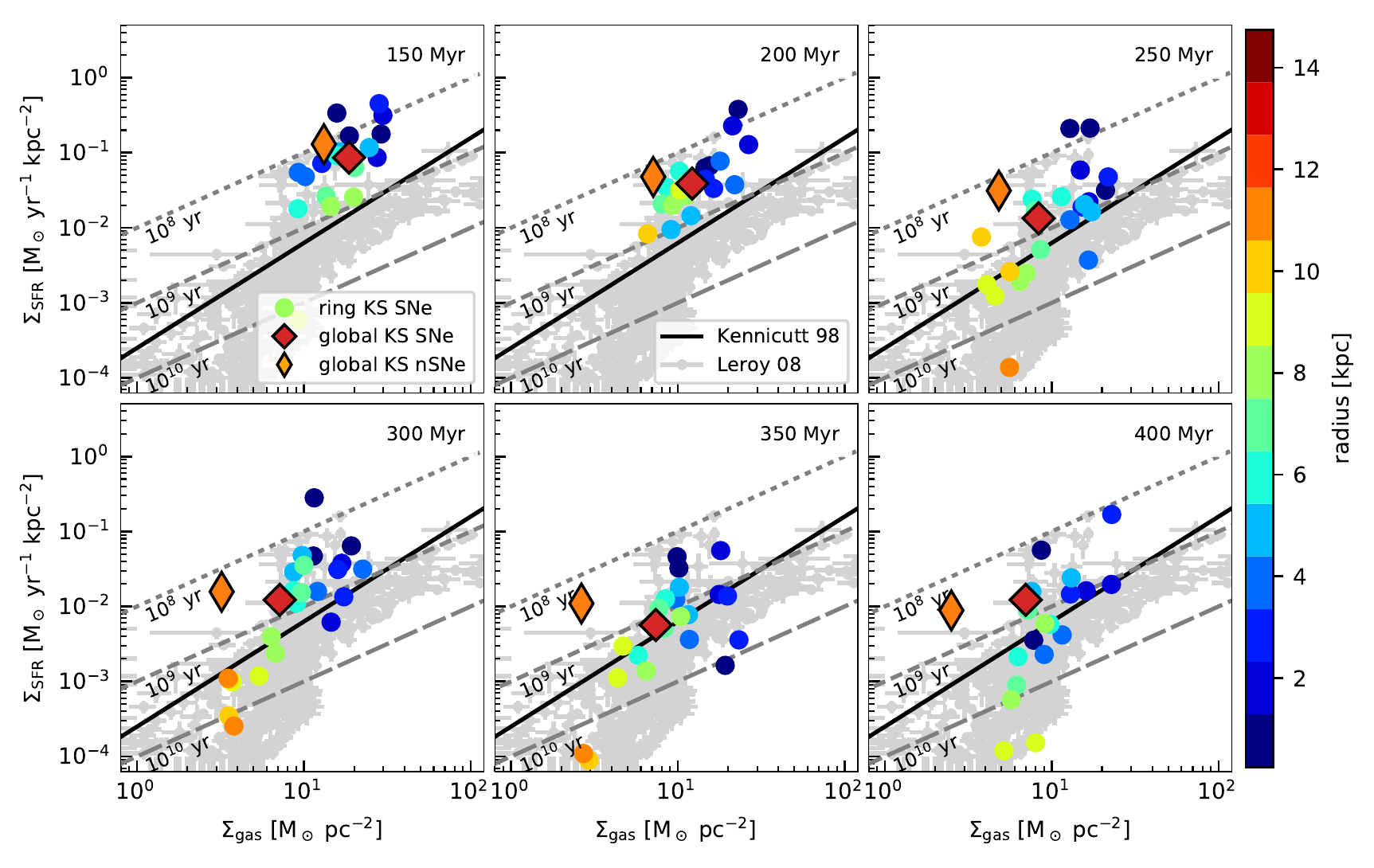}
\caption{KS Relation for different rings (indicated by different colors) within
the galaxy at a given time. The global KS for the \nSNe/\SNe simulation is
calculated within a fixed radius of $r=8$~kpc. The solid black line is the KS
relation.  The different dotted lines show constant depletion times of 10, 1,
and 0.1~Gyr from bottom to top. We compare the simulated values to
observational data of Leroy et al. (2008).  Our \SNe{} simulation compares well
with the observational sample. The rings as well as the global value start
within the star-bursty regime above the KS value and move closer around the KS
relation with time. SNe regulate SFR early for low surface density regions,
whereas for higher surface densities (above solar neighborhood) it takes a
little bit longer, potentially because SNe do not act fast enough. Without SNe
the global KS value stays well within the star-burst regime and/or outside
observational values.}
\label{fig:surfSFRringsTime}
\end{figure*}

After around $\sim$~300~Myr our \SNe{} simulation compares well with the KS relation within the
observational uncertainties for normal star-forming
galaxies in all rings. For higher gas surface densities or within the inner
7~kpc in the disc the SFR surface density is at the upper end of the margin or
more in the star-bursty regime (see also Figure~\ref{fig:surfSFRringsTime} and
discussion below). The SFR surface densities are, for rings within 7~kpc, still
declining at the end of the simulation, whereas the gas surface densities
(indirectly shown via the grey dotted line) remains fairly constant after
$\sim$~250~Myr. In other words, we see that these high density regions have almost 
but not yet completely reached a self-regulated steady state. However, for gas 
surface densities close to the radius of our
solar neighborhood (between 7-9~kpc from the centre of the galaxy) we find that
$\Sigma_{\mathrm{SFR}}$ lies well within the KS relation after around
$\sim$~250~Myr of simulation time. Additionally, both the gas and SFR surface
densities stay after $\sim$~250~Myr constant over time, showing that this
region within the galaxy is due to the presence of SNe feedback self-regulated.
Finally, for lower \sigmaGas{} at radii larger than 9~kpc the SFR surface
density lies below the KS but still within the observational uncertainty
plotted. At the same time both \sigmaSFR{} and \sigmaGas{} are still declining.
As already seen in Figure~\ref{fig:GasTemperatureMapComparison} the gas
structure outside $\sim$~9~pc is at the end of the simulation still very smooth
and cold. It is therefore not surprising that this region has not yet reached a
self-regulated state.   

Figure~\ref{fig:surfSFRringsTime} shows the relation between SFR and gas
surface densities for different rings (indicated by different colors) within
the galaxy at different times. We also show the global KS for the \nSNe/\SNe{}
simulation computed from the total SFR rate and the total gas mass within
$r=8$~kpc.  To guide the eye we also plotted the KS relation
\citep{Kennicutt98} as a solid black line. The different dotted lines show
constant depletion times of 10, 1, and 0.1~Gyr from bottom to top. The obtained
simulation relations are compared to the observational data of
\cite{Leroy+2008} that also show significant scatter around the KS value.

Similar to the findings from Figure~\ref{fig:surfSFRrings} we find that all the
rings, as well as the global KS values, compare well with the observational
sample shown. At the beginning of the simulation ($t < 250$~Myr)
$\Sigma_{\mathrm{SFR}}$ lies, for the rings as well as the global value, above
the KS relation and thus within the star-bursty regime. Around $t=250$~Myr the
simulation scatters close around the KS relation.  The rings with higher
surface densities or, in other words, the rings closer to the centre of the
galaxy lie more above the KS relation and approach the KS relation around
300~Myr. The rings with lower surface densities are already early on close to
the KS relation. The global KS value for the \SNe{} simulation shows that gas
is converted  into stars on a typical time scale of 1-2~Gyr, which is much
longer than the free fall time. The resulting star formation rate in our simulation
is thus rather inefficient. Without SNe regulation the global value stays
well within the star-burst regime and/or outside observational value. This is
in agreement with the discussion above. 

To conclude, we find that in our simulations SNe alone are efficient enough to
regulate star formation within solar neighborhood surface densities and lower.
Within the solar neightborhood the galaxy reached a self-regulated
steady-state. This is not the case for the outskirts of the galaxy (radii
larger than 9~kpc) where the gas has not yet collapsed and is still smooth.
For larger surface densities it takes a little longer for the SNe to regulate
the rate at which stars form. Additionally, these higher density regions never
reach a self-regulated steady-state state as \sigmaSFR still keeps decreasing
at the end of the simulation. Early feedback by, for instance, radiation from
the stars or stellar winds might help to limit SFR as it prevents gas from
accreting onto the stellar clusters (Bieri et. al in prep). 
\subsection{Multi-phase ISM Structure}
\label{sec:ISM}
\begin{figure}
\plotone{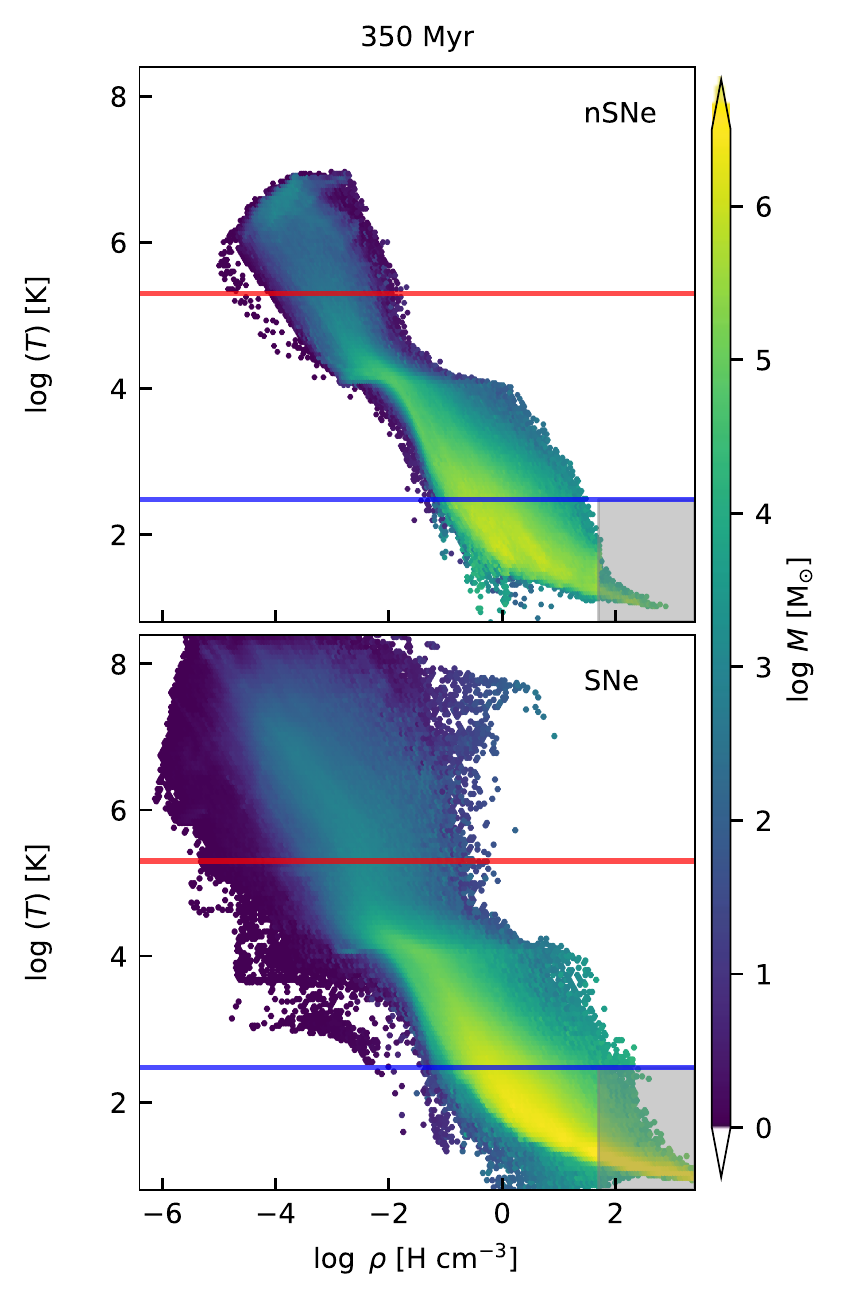}
\caption{Two-dimensional temperature versus density diagram of all the gas
cells within a cylindrical region of $r\leq 10$~kpc and  $z \leq \pm 250$~pc.
We average the phase diagram over the time period of $\pm$~10~Myr around
350~Myr. The diagram is colored by the total mass within the pixel. The lines
highlight the temperature cuts for the different gas phases in blue (cut for
cold/warm gas) and red (cut for warm/hot phase). We additionally highlight the
gas that is above the sink formation threshold (i.e
$\rho_{sink}$=50~H~cm$^{-3}$).  Due to the SNe feedback a lot more gas
populates regions that are thermally unstable because the explosions disrupt
the gas and eject hot gas from within the star forming disc. The gas within the
\SNe{} simulation reaches both lower and higher densities.}
\label{fig:PhaseDiagram}
\end{figure} 
\begin{figure*}
\plottwoH{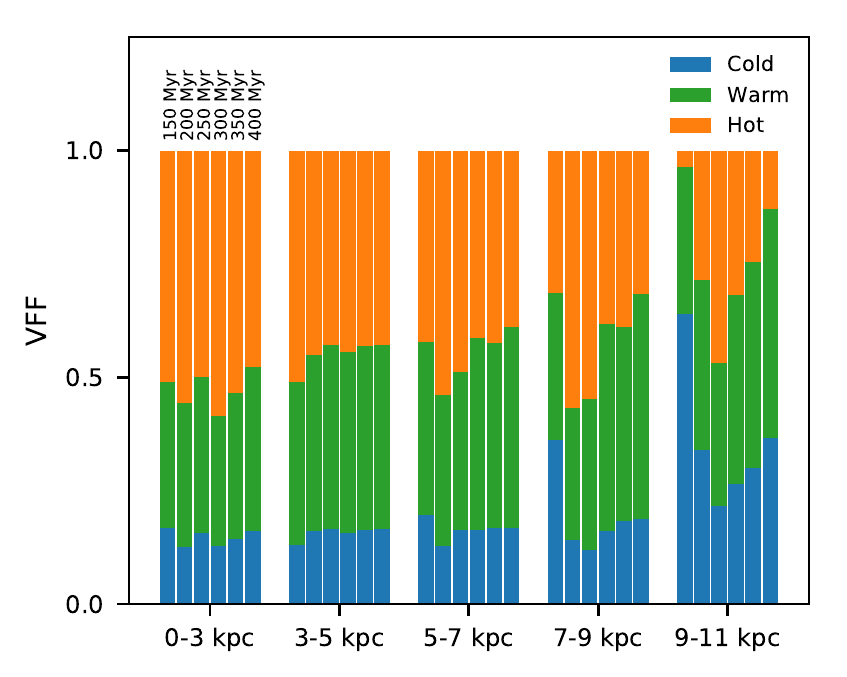}{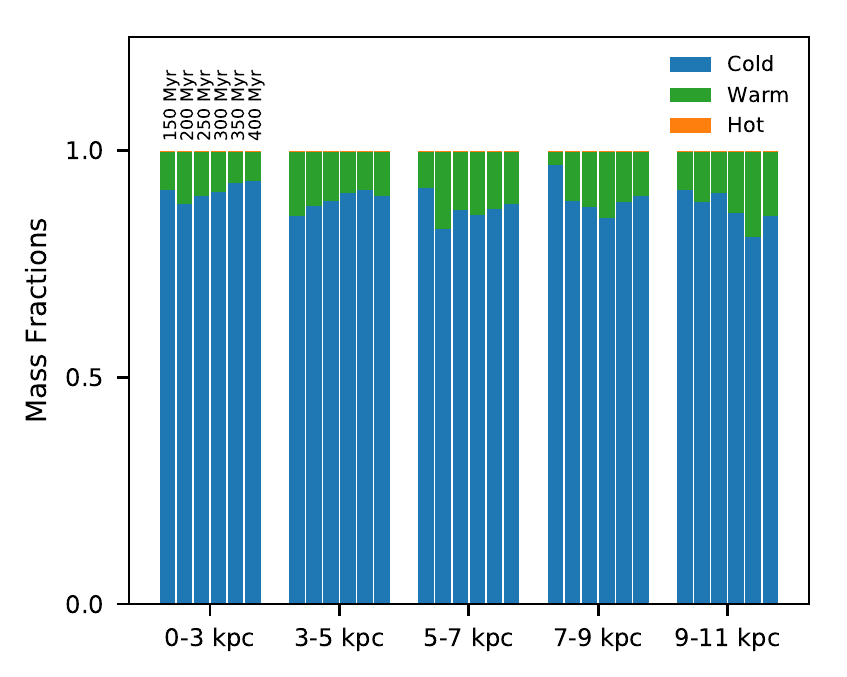}
\caption{Volume Filling Fractions (VFFs) and Mass Filling Fractions (MFFs) for
different times (150--400~Myr in 50~Myr spacing) and different cylindrical
rings (radii indicated in the axis label) for the \SNe{} simulation at
the top/bottom. For the \SNe{} simulation the hot gas fills most of the volume
($\sim$~50\%), followed by the warm ($\sim$~30-35\%), and then cold phase
($\sim$~15-20\%). Once the SFR surface density reaches KS value for the
corresponding surface densities the VFF is stable with time, except for the
regions further out in the galaxy ($r>9$~kpc). The high density cold gas makes
up most of the mass in the galaxy ($\sim$~85-90\%), followed by lower density,
warm gas ($\sim$~10-15\%) produced mostly through SN heating. The hot phase
only makes up to 0.5\%. Comparing to models for the ISM phase based on 
observations we underestimate/overestimate the VFF in the warm/hot phase. Early feedback by radiation that disrupts the dense gas in star-forming regions and thus
increases/decreases the VFF in the warm/hot phase would likely help.}
\label{fig:VFFMFF}
\end{figure*} 
\begin{figure*}
\plottwoH{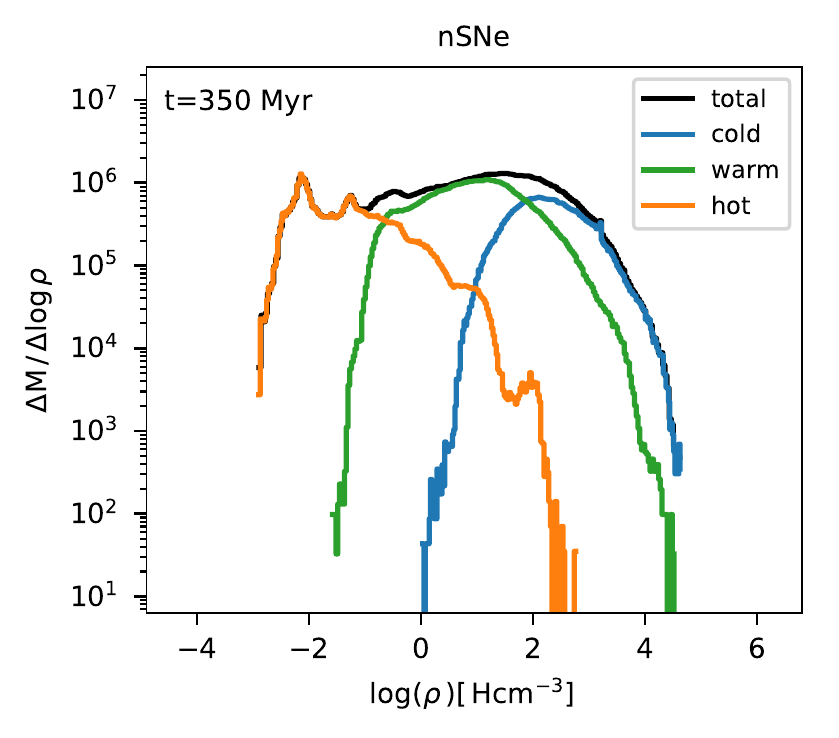}{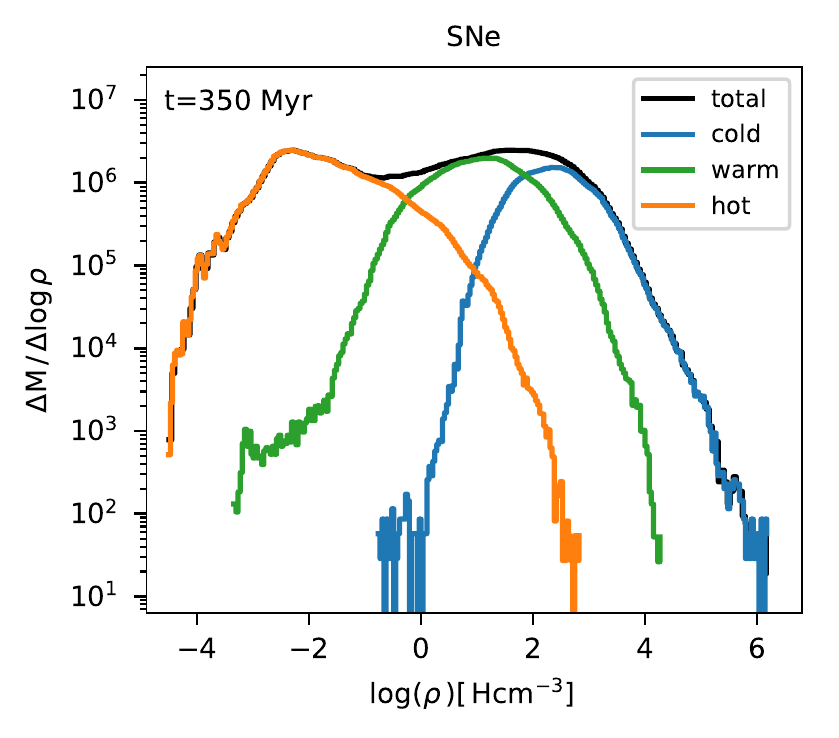}
\caption{Differential gas mass distribution as a function of the gas density at
time $t=350 \pm 10$~Myr for the simulated \nSNe/\SNe~galaxies.  The probability
density function (PDF) is calculated using only gas within a cylindrical region
of $r=10$~kpc and z~$\leq \pm 250$~pc. The PDF shows the total distribution
(black), the cold (blue), warm (orange), and hot (red) gas phase within the
ISM. As a consequence of SNe feedback a multi-phase structure forms in the ISM.
This leads to a bimodal character of the mass distribution for the \SNe{}
simulation. The gas within the \SNe{} simulation reaches both lower and higher
densities as SNe explosions halt the cold gas from further collapse, and heat
and disperse the gas.}
\label{fig:DifferentialGas}
\end{figure*} 
\begin{figure}
\plotone{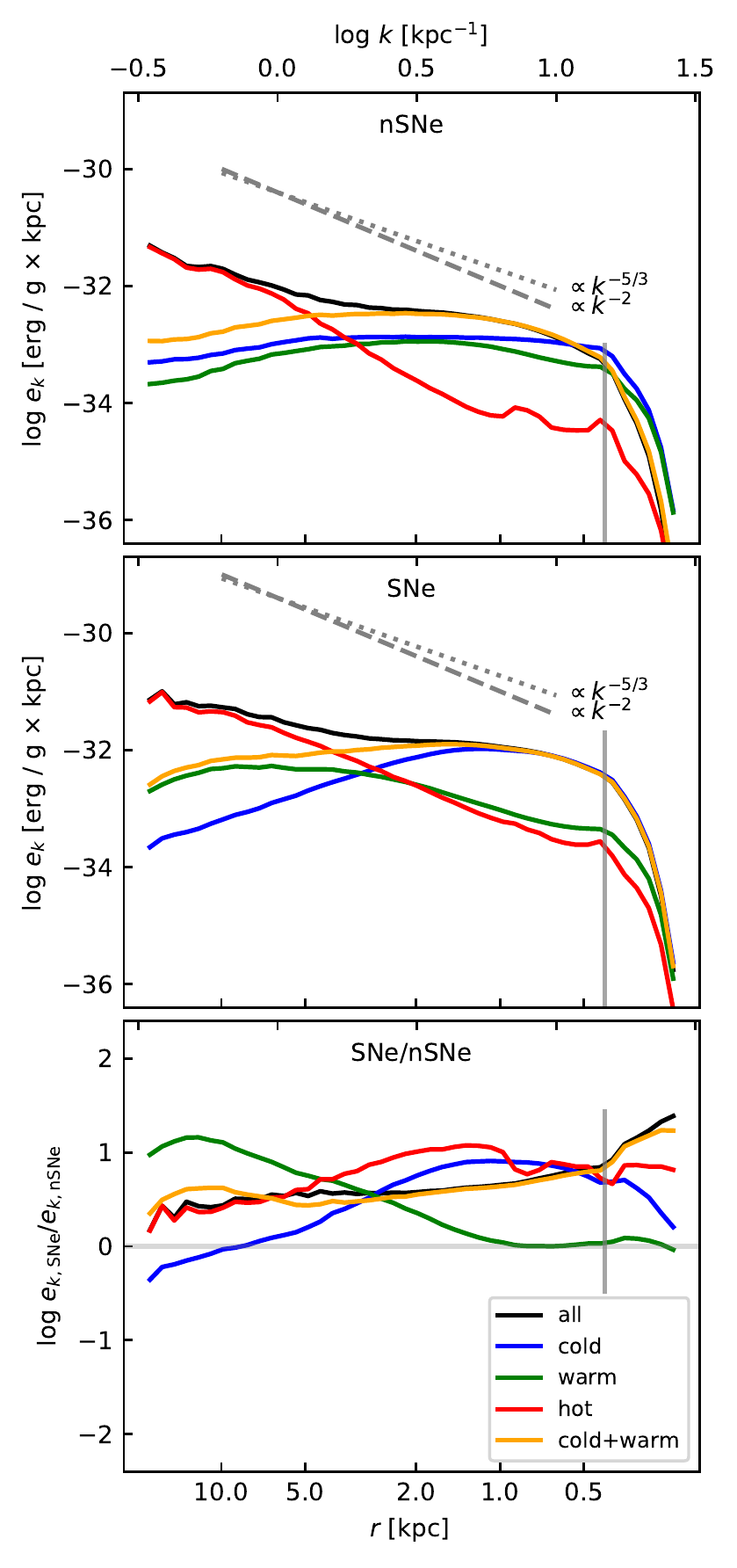}
\caption{Specific kinetic energy power spectrum at $t=350$~Myr averaged over
$\pm$~10~Myr. The top/middle panel show the different contribution of the cold
(blue), warm (orange), and hot (red) phase to the total power spectrum (black)
of the \nSNe/\SNe{} simulations. The bottom panel shows the ratio of the
specific kinetic energy power spectra of \SNe{} over \nSNe{}.  The physical
scale $l$ is connected to the wave vecotr via $k = 2 \pi / l$. The vertical
line corresponds to a physical length of 420~pc ($\sim 46 \Delta x$) and is also 
where the power spectra drop of steeply. The power spectra of the
\SNe{} simulation have more specific kinetic energy at all scales for all gas
phases. SNe feedback is thus an important driver of the density and energy
structure of the ISM. For the hot gas turbulence is clearly also driven by a combination of gravity and shear as both simulations show a power-law behavior.  Only the power spectrum for the \SNe{} simulation shows super-sonic ($e_k \propto
k^{-2}$, between 550~pc and 10~kpc) for the hot and Kolmogorov ($e_k \propto
k^{-5/3}$, between 450~pc and 4~kpc) scaling for the warm gas.  Finally, the
cold gas does not show any turbulent scaling as the SNe feedback does not
manage to disrupt the cold and dense clouds within the galaxy.}
\label{fig:PowerSpectrum}
\end{figure} 
Capturing a turbulent multi-phase ISM is an important goal of the \satin{} simulation project. To analyse the properties of the ISM structure we define
three different gas phases in our simulation: The cold (T < 300~K), warm (300~K <= T < 2 $\times 10^4$~K), and hot (T >= $2 \times 10^4$~K) medium. The properties of these phases is regulated by a complex interplay between cooling, self-gravity, turbulence, shear, and heating from SNe feedback during the evolution of the galaxy. In this Section we show that with SNe feedback included the gas within the galaxy is multi-phase, the phase diagram populates larger regions that are thermally unstable, and that after some time the VFFs as well as MFFs  stay remarkably stable within 7~kpc. 

Figure~\ref{fig:PhaseDiagram} shows the phase diagram of all the gas cells within
a cylindrical region of $r\leq 10$~kpc and  $z \leq \pm 250$~pc, averaged over
the time period of $\pm$~10~Myr around 350~Myr. The two-dimensional temperature
versus density histogram is additionally colored by the total mass within the
pixel. We highlight the temperature cuts for the different gas phases in blue
(i.e temperature cut for cold gas) and red (i.e temperature cut for hot gas).

For both \SNe{} and \nSNe{} simulations the density covers several orders of magnitude
in the range from $\sim
5 \times10^{-5}$ to $\sim 5 \times 10^2$~H~cm$^{-3}$.  Both simulations have a
large fraction of warm gas that is close to the equilibrium cooling curve. Due to 
SNe feedback, however, a lot more gas populates regions in the phase 
diagram that are thermally unstable as the explosions disrupt the gas and eject hot gas
from within the star forming disc. As a result the hot gas covers a larger
density range for the \SNe{} simulation.  The cool gas reaches densities of
$\sim 10^{-1}$~H~cm$^{-3}$ as only gas that is dense enough can effectively
self-shield from the UV background, cool down, and eventually become dense
enough to form stars.  The coloration of the phase diagram indicates that the cold, 
high density gas makes up most of the mass within the galaxy.  

We highlight the gas that is
above the sink formation threshold (i.e $\rho_{sink}$=50~H~cm$^{-3}$). Star
clusters can only form from gas with this density and higher, but only after it
will collapse gravitationally (see Section~\ref{sec:SinkFormation}). For the
\SNe{} simulation there is a substantial amount of colder (below 10~K) high
density gas above the sink formation threshold. This is because the SNe disrupt
the high density gas more and as a result fewer high density gas clumps fulfil
the requirements for sink formation/accretion. However, we also possibly miss important
physical processes such as cosmic rays, ionisation, and winds from the
stars. Those effects could likely prevent the formation of the coldest gas in
this diagram \citep[see also][]{Geen+2015b, Peters+2015, Girichidis+2016,
Simpson+2016, Gatto2017, Peters+2017}. We plan a more detailed investigation in
future work.

We identify the gas volume and mass in the different phases within cylindrical
rings (radii indicated in the axis label) and z~$\leq \pm 250$~pc for different
times (150--400~Myr in 50~Myr spacing) in Figure~\ref{fig:VFFMFF}.  Generally,
after 150~Myr the hot gas fills most of the volume ($\sim$~50\%), followed by
the warm ($\sim$~30-35\%) and then cold phase ($\sim$~15-20\%). The volume
filling fractions in the central region ($r<7$~kpc) stay remarkably
stable with time. Around the solar neighbourhood ($r=7-9$~kpc) the VFF
oscillates more. However, once the SFR surface density reaches the KS value
for these gas surface densities at around 300~Myr (see
Figure~\ref{fig:surfSFRrings}), the VFF also shows less variation over time.
For regions further out in the galaxy ($r>9$~kpc), where stars start to form
much later and the density region is still very smooth and cold at the end of
the simulation time, the VFF continues to vary for the duration of our
simulation. 

As already noted above, cold, high density gas makes up most of the mass in the
galaxy's disk ($\sim$~85-90\%), followed by lower density, warm gas
($\sim$~10-15\%) produced mostly through SN heating. The hot phase only makes
up to 0.5\%. For regions within 9~kpc the MFFs stay, after $\sim$~250~Myr,
stable with time, whereas it keeps varying for the outer region of the disc. 

This solidifies our findings (see Section~\ref{sec:SFR-KS}) that within the
solar neighborhood the galaxy reached, after 300~Myr, a self-regulated
steady-state. Larger surface density regions (radii smaller than 7~kpc) have,
due to \sigmaSFR{} still decreasing, not yet reached a steady-state although
\sigmaGas{}, VFF, and MFF stay remarkably constant over time. This is because
most of the gas is in this region in the dense cold star forming clouds where
SNe feedback slightly disrupt, but not destroy, the clumps such that the gas
does not fulfill the requirement for sink formation/accretion.  The outskirts
of the galaxy also have not reached a steady-state as the gas within this
region has not yet formed a stable multi-phase structure causing the VFFs,
MFFs, as well as \sigmaSFR{} and \sigmaGas{} to not be constant with time.  

The same analysis for the \nSNe{} simulation (not shown) reveals that the inner region shows a similar behaviour as for the \SNe{} simulation, indicating that the 
evolution for this inner region is mostly driven by rotation and shear and not SNe. The intermediate regions ($r = 3-9$~kpc) have, for the \nSNe{} simulation, not reached a steady state as the mass and volume filling fractions for the cold phase keeps decreasing for the length of the simulation due to the ever accreting sinks. The outskirts of the galaxy equally has not reached a steady-state for similar reasons as for the \SNe{} simulation. The \nSNe{} simulation has most of the mass in the cold phase and close to no mass in the hot phase\footnote{The \nSNe{} simulation has no process to create hot gas. All of the hot gas comes from accreted hot gas from the CGM.}.

As we will discuss in more detail in Section~\ref{sec:Discussion} compared to
observations the \satinI{} \SNe{} simulation matches the VFF of the cold phase range, but slightly underestimate/overestimate the VFF in the warm/hot phase. \cite{Peters+2017} reported similar results for their SNe only simulation and showed that only their simulation including radiation from the stars approached observed values. This early pre-SN feedback via energetic radiation of massive stars disrupts the dense gas in star-forming regions and by doing so increases/decreases the VFF in the warm/hot phase. Simultaneously, this also may help to self-regulate SFR in the central region. 

The mass-weighted gas density probability density function (PDF) displays each
gas phase in the ISM in another way. Figure~\ref{fig:DifferentialGas} shows the
PDF for the \nSNe/\SNe{} simulation at 350~Myr averaged over $\pm$~10~Myr. To
calculate the PDF we only take the gas within the disc using a spatial cut of
$r=10$~kpc and z~$\leq \pm 250$~pc (as for the phase diagram above)\footnote{We
tested several different radii and height cutoff values to confirm that the
qualitative result does not change}. The different colors in the Figure
highlight the contributions of the cold (blue), warm (orange), and hot (red) to
the total (black) gas within the ISM. As a consequence of the SNe feedback, the
galaxy shows a clear bimodal character in the total mass distribution. This
multi-phase ISM character, especially the prominent low density peak, is less
apparent for the \nSNe{} simulation. 

The warm and cold dominated phases transition at similar densities
($\sim$~100~H~cm$^{-3}$) for both the \nSNe{} and \SNe{} simulation. At higher
densities cold gas dominates in both distributions. The gas within the \SNe{}
simulation reaches both lower and higher densities compared to the \nSNe{}
simulation. The injected kinetic energy due to the SNe explosions prevents some of the cold gas from accreting onto the sink leading to the observed higher densities. Additionally, the SNe also heat and disperse the gas causing the distribution to have lower densities, and the hot phase to dominate to higher densities ($\sim 10^{1.8}$~H~cm$^{-3}$) for the \SNe{} simulation compared to the \nSNe{} simulation ($\sim 10^{1.2}$~H~cm$^{-3}$). 

To summarise, the \satinI{} simulations capture a distinct multi-phase ISM with
three coexisting gas phases. With SNe included the gas populates larger regions
in the phase diagram especially some that are thermally unstable.  Because of
the lack of SNe the ISM the \nSNe{} is smoother and does not reach as
low/high densities.  As we will also show more in the next
Section~\ref{sec:Turbulence}, the creation of the multi-phase ISM is, in our
simulated galaxies, not only due to SNe feedback but also cooling, gravity, and
shear. 
\subsection{ISM Turbulence}
\label{sec:Turbulence}
In this Section we will look more closely at the impact of SN feedback in
shaping the energy structure of the ISM.  The turbulent ISM can be quantified
in Fourier space using power spectra, which is often used in the context of
turbulent flows.  On galactic scales, simulations
\citep[e.g.][]{Wada+2002,Agertz+2015, Grisdale+2017} and observations of the
neutral ISM in nearby galaxies \citep[e.g.][]{Stanimirovic+1999, Bournaud+2010,
Combes+2012, Zhang+2012, Dutta+2013}  show that turbulent scalings are present
over several orders of magnitude in scale. It is, however, not yet fully
established what physical mechanisms (gravity, shear, accretion, feedback)
maintain the turbulence on these galactic scales. The interested reader may be referred to \citep{Pfrommer+2022} investigating in more detail different turbulence driving mechanisms in MW-like discs.

We calculate power spectra of the specific kinetic energy by taking the absolute
square of the Fourier transform of $\sqrt{\rho} \textbf{v}$ and dividing in the end
the kinetic energy power spectra by the total gas mass within the box. The
chosen uniform box is centred on the disc, has a physical extent of
$r_{\mathrm{box}} \simeq 15$~kpc, is zero padded to $2 \times
r_{\mathrm{box}}$, and has a uniform resolution of 40~pc.  A more detailed
description of the power spectra calculation can be found in
Appendix~\ref{sec:PowerSpectrumCalculations}. 

Figure~\ref{fig:PowerSpectrum} shows the kinetic energy power spectrum at
$t=350$~Myr averaged over $\pm$~10~Myr. In the top and middle panel we show the
different contributions of the cold (blue), warm (orange), and hot (red) phase
to the total power spectrum (black) of the \nSNe{} and \SNe{} simulation.  In the
bottom panel the \SNe{} over \nSNe{} ratio of the power spectra is shown. The vertical line shows the wavenumber corresponding to a physical length of 420~pc ($\sim 46 \Delta x$) and is where the power spectrum drops of steeply. 

The kinetic power spectra for the two simulations show that at large scales
most of the kinetic energy is in the hot gas. The power-law behavior of the hot
gas shows that it is also turbulent and thus most of the energy resides at the
top of the self-similar cascade where turbulence is driven. Both \nSNe{} and \SNe{} simulations show a power-law behavior in the hot phase. This indicates that turbulence is in this phase not only maintained by SNe but also other physical mechanisms such as gravity and shear. However, the scalings are only for the hot
gas in the \SNe{} simulation, between 550~pc and 10~kpc, in good agreement with that for super-sonic turbulence (i.e, $e_k \propto k^{-2}$, \citealp[][]{Burgers1948}).

At 4~kpc the total power spectra transitions into a shallower relation and the
kinetic energy of both the cold and warm gas together start to dominate. We
find that only the slope of the warm gas for the \SNe{} simulation follows that
for subsonic Kolmogorov turbulence ($e_k \propto k^{-5/3}$, between 450~pc and
4~kpc; \citealp{Kolmogorov1991}). The cool gas is not turbulent in both \nSNe{}
and \SNe{} simulations because the corresponding power spectrum does not 
show a power-law behaviour. The cool gas is, however, still important for 
the energetics of the ISM, especially in the \SNe{} simulation, on scales smaller than 3~kpc. It seems that for the cold gas, large scale turbulence (i.e large scale rotation and gravitational instabilities) is no longer present, as most of the power is now `locked-up' in the dense cold star forming clouds instead of being disrupted and returned back to the large scale driving. This is because the dense star
forming clouds are only slightly disrupted, but not destroyed, by the SNe explosions. For all gas phases and both simulations the power spectrum drops off steeply below 420~pc ($\sim 46 \Delta x$). The ratio of the power spectra for the two simulations shows that the \SNe{} simulation has more kinetic energy at all scales and for all gas phases. This shows again that SNe increase the kinetic energy of the cold gas that is cooling out of the warm phase. We also find that feedback regulation results in steeper power spectra for the warm gas (similar results have been found by \citealp{Grisdale+2017}). We find the same quantitative result for the kinetic energy power spectra. 

To conclude, by looking at the power spectrum we find that SN feedback is an
important driver of the density and energy structure of the ISM (up to
$\sim$~12~kpc) and that it helps to shape the power spectra, especially for the
warm gas phase.  However, the comparison between the \SNe{} and \nSNe{}
simulations shows that gravity, shear, and accretion also contribute
significantly to the specific kinetic energy at all scales and are therefore
also important for the energetic structure of the ISM. In our simulations most of the 
power in the cold phase is in the star forming clouds as SN feedback does
not manage to fully disrupt the dense clumps and to drive turbulence
on galactic scales for the cold gas. The effect of early stellar
feedback by radiation and winds as well as cosmic ray heating would prevent the
formation of the coldest gas within the galaxy and thus would change the
energetics of the ISM.  We leave the detailed investigation of the missing
physical mechanisms and their effect on the ISM and power spectrum to future
work. 
\subsection{Gas Dynamics and Flows}
\label{sec:gasDynamicsFlows}
Gas flows in and out of galaxies are important in regulating star formation as
well as to drive turbulence within the galaxy. In this Section we will discuss
that the SNe generate multi-phase gaseous outflows. The cold and warm gas is
recycled and builds a fountain flow sustaining the formation of stars. Later the
SNe push the hot gas far enough from the galaxy such that it does not fall
back. We will only discuss the \SNe{} simulation in this Section as the \nSNe{} 
simulation does not create any outflow. 

\begin{figure}
\plotone{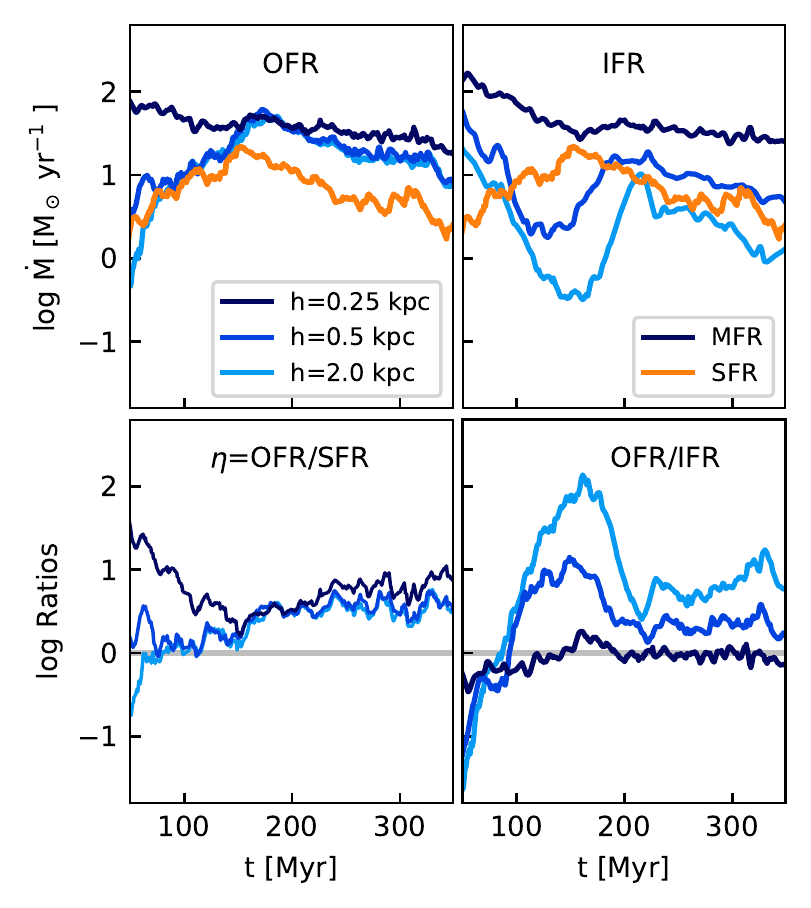}
\caption{Total outflow rate (OFR; top panel) and inflow rate (IFR; middle
panel) as a function of time; calculated at different heights and compared
against the total SFR. Bottom panel shows the mass loading ($\eta$=OFR/SFR) as
well as OFR/IFR. The mass flux is calculated at $r=13$~kpc and within a slab at
different heights (h=0.25, 0.5, 1, 2~kpc). The IFR close to the disc follows
closely the OFR at the same height. The ratio of OFR/IFR is above unity the
further away from the galaxy.  After 200~Myr the mass loading factor evolves
around the same value ($\sim$~0.1) for all heights. After 200~Myr the SNe
explosions are efficiently pushing the gas far away from the galaxy.}
\label{fig:OFR_IFR_SFR}
\end{figure}

Figure~\ref{fig:OFR_IFR_SFR} shows the measured gas outflow rate (OFR; top left
panel), inflow rate (IFR; top right panel), mass loading ($\eta$=OFR/SFR;
bottom left panel), as well as the ratio of the OFR over IFR (bottom right
panel); all measured at different heights (h=0.25, 1, 2~kpc) also 
indicated in the legend. Overplotted
is the total SFR within the galaxy for comparison (top/middle panel). 

To compute the mass flux we only consider gas at a radius of $r \le 13$~kpc and
within a slab at different heights measured from the midplane of
the galaxy\footnote{We used this approach to compare with other simulation
results such as Marinacci 2019.}.  The gas mass flux is then calculated as  
\begin{equation}
\dot{M}_{\rm gas} = \oiint \rho \,\vec{v} \cdot \hat{\vec{r}}\, \mathrm{d}S
= \sum_{i \in \mathrm{slab}} m_{i}\, \vec{v}_{z,i} \, / \Delta z ,
\label{massflux}
\end{equation}
where $i$ denotes the index of a cell within the slab. We adopt a slab
thickness of $\Delta z$=500~pc. Positive/negative velocities $v_z$ with
respect to the disc plane contribute to the outflow/inflow. 

Until the SFR declines around 200~Myr the OFR is highest closer to the galaxy
and then progressively decreases further away from the galaxy. Afterwards the
OFR settles around 30~M$_\odot$/yr at all heights, with only a slow decline over time. This
is half an order of magnitude higher than the measured total SFR within the
galaxy, also shown in the mass loading factor ($\eta=\sim 3-10$) in the bottom
panel. The OFR, calculated at $h \ge 1$~kpc, noticeable follows, with a short
delay, the SFR.  After 200~Myr the mass loading factor settles at the same
value for all heights ($\eta=\sim 7$) indicating that, after some time, the outflow is directly
correlated with the SFR. But not only the global OFR is correlated with the
SFR, as similar outflow measurements for different radii within the disc also
have shown that the OFR is strongest near the galactic centre, where star
formation is most effective.  

The IFR decreases for the first 150~Myr and more drastically the further away
in height from the galaxy. This is most likely due to adiabatic contraction and further
settling of the initial conditions. With the increasing SFR the IFR also starts
to increase with a short time delay at distances above 1~kpc. Once the SFR
decreases, around 200~Myr, the IFR also starts to decrease, again with a short
time delay that is longer at larger galactic heights. 
The highest IFR is closest to the disc and one measures  progressively lower rates
the further out vertically from the galaxy.  The IFR close to the disc follows closely the
OFR calculated at the same height; also shown by the very close to unity ratio
of the OFR over IFR shown in the bottom right panel. The OFR/IFR ratio is further
above unity at larger galactic heights showing that some of the gas that
is pushed away more than 2~kpc above the galaxy also likely does not fall back. 

\begin{figure}
\plotone{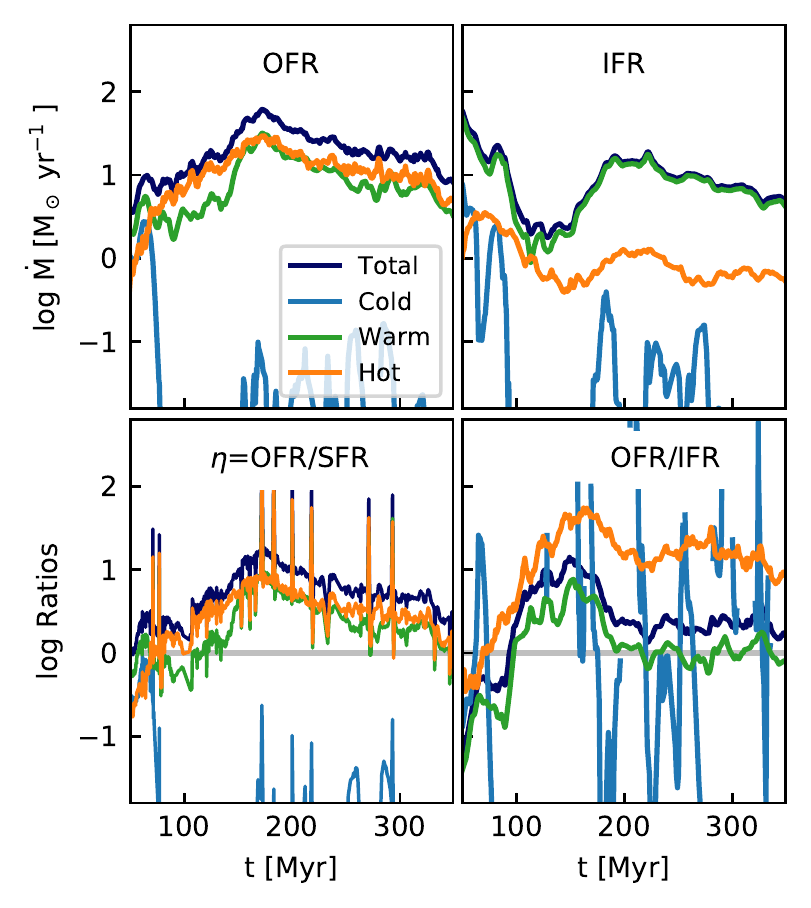}
\caption{OFR/IFR as a function of time for the different phases in the
top/middle panel, both also compared against the total SFR. Mass loading
($\eta$=OFR/SFR) and OFR/IFR for different phases is shown in the middle panel.
After $\sim$~150~Myr the warm gas that is pushed out by the SNe has the highest
contribution to the total mass outflow/inflow rate. The hot gas contributes
little to the total OFR. However, most of the gas that leaves the galaxy is hot
as it has the highest OFR over IFR ratio. The warm gas builds a fountain flow
and falls back onto the galaxy fueling star formation.}
\label{fig:OFR_IFR_phases}
\end{figure}

At the beginning, most of the gas that is pushed to such distances is hot,
followed by the warm and cold gas (see Figure~\ref{fig:OFR_IFR_phases}).  Once
the SFR decreases after $\sim$~200~Myr the hot gas, which is mostly the SNe
ejecta itself, and the warm gas contribute in similar fractions to the total
ouflow rate.  Some of the outflowing warm gas also originates from
cooled (adiabatically and radiatively) gas that was previously heated by SNe.
The vast majority of the inflowing gas is warm, followed by the hot and then
cold gas. The ratio of OFR/IFR is higher for the hot than warm gas and smallest
for the cold gas. This shows that most of the gas that leaves the galaxy is
hot, whereas most of the warm gas falls back even though it reaches as far as
1~kpc away from the galaxy.  The cold gas does not reach such large distances
away from the galaxy and thus contributes little to the total outflow and
inflow mass at this distance. Similar measurements closer
to the galaxy ($h=0.25-0.5$~kpc) show that cold gas gets entrained in the outflows up to these distances. It then, however, quickly falls back onto the galaxy.

The mass flow measurements show that the SNe ejecta are at early times not
strong enough to push the gas to large distances and instead immediately
fall back onto the galaxy. Later, the SNe push mainly hot ejecta gas far
enough from the galaxy such that it does not fall back. The warm and cold gas
falls mostly back onto the galaxy, with the warm gas reaching higher distances
away from the galaxy. The ejected gas in our simulations is not lost to the
galaxy but rather builds a fountain flow where the accreted gas fuels star
formation.  Such a cycle is important for the metallicity and angular momentum
build up of the disc and hot corona
\citep[e.g.][]{Marinacci+2010,Uebler+2014,Christensen+2016,Angles-Alcazar+2017}. As we will discuss in more detail in the next Section~\ref{sec:Discussion} the
obtained outflow rates as well as the fountain flow character of the gas in our
simulations are comparable to similar galaxy scale simulations. 
%
\section{Discussion}\label{sec:discussion} 
\label{sec:Discussion}
In this paper we present simulations of a MW-like disc galaxy with a
turbulent and multi-phase ISM structure run with the stellar formation and SNe
feedback model of the \satin{} project. The goal
of this project is to bridge the gap between large and small scales to
understand the interaction of AGN feedback with the multi-phase gas and how
large-scale winds are driven. Because the ISM properties are so tightly linked
to star formation and stellar physics we also incorporate a
stellar feedback model that regulates star formation as well as the
properties of the ISM. For this, we model the formation of star clusters and
their individual stellar population with sink particles and follow the
evolution of the massive stars formed within the cluster to get realistic SN
delay times. The SN implementation in the \satin{} model adaptively adjusts to
the local environment and switches between the injection of thermal energy and
momentum depending on the surroundings. In this paper we discussed in detail
the effect of SNe feedback with individual delay time distributions on the 
multi-phase and turbulent ISM as well as compared the results with observations.
The inclusion of other stellar feedback
processes from massive stars such as radiation and stellar winds is work in
progress (Bieri et. al, in prep.). 

Other simulations in the literature aim to bridge the gap between
large-scale models and more detailed ISM simulations for MW-like environments
\citep[e.g.,][]{Li+2005, Li+2006, Agertz+2013, Rosdahl+2015, Grisdale+2017, Hopkins+2018,
Martizzi+2019, Marinacci+2019, Tress+2020} and there is a vast body of work studying the solar neighbourhood ISM and the subsequent outflow properties \citep[see e.g.][]{Kim+Ostriker2015, Martizzi+2015, Walch+2015}. 
In this Section we will discuss the results of the \satinI{} simulations 
in the context of other full galaxy-scale simulations of MW-like galaxies as well as compare with more detailed ISM studies.  Our \satin{} model is very similar as used for the stratified solar neighbourhood-like simulations of the SILCC project \citep{Gatto2017, Peters+2017, Rathjen+2021}. Given the comparable setup, this is a great opportunity to compare our results from a full galaxy wide simulation with higher resolution simulations. This will allow us to investigate the need for different stellar feedback channels in light of our results and to determine the need for a more realistic galaxy-scale environment. 

We like to emphasise that one novelty of the \satin{} model is that it adopts 
sink particles to model the formation of star clusters as well as that it tracks the
evolutionary state of single massive stars to get realistic SNe delay times. The only other full galaxy-scale simulations of MW-like galaxies with a similar sink particle approach are \citet[][]{Li+2005, Li+2006} and \citet[][]{Tress+2020}. The former present
various disc simulations with a gas resolution of $\sim 5-10 \times 10^4$~\Msun (versus $5 \times 10^3$~\Msun for our simulations) for their two MW-like disc simulations. They 
also adopted a different star formation model and convert the mass of sink particles to stars using a fixed star formation efficiency. Additionally, they only include 
stellar feedback implicitly by maintaining a constant gas sound speed. The simulations
of \citet{Tress+2020} use a very similar sink particle as well as
an individual SNe delay time distribution approach (although the exact implementation and treating of the sink particles differ slightly). Because, however, a sink particle approach is not the only possible way of simulating star formation within high-resolution MW-like galaxies we will additionally include in our discussion other simulations of MW-like galaxies using a different approach for star formation and stellar feedback as well as using delay times calculated with the assumption of average
stellar populations. 

We find that in \satinI{} SNe, gravity, and shear manage to keep global star formation within observed scaling relations and move the galaxy from the star burst regime above the KS value closer to the KS relation. Interestingly, very similar results have been found by \citet{Li+2005, Li+2006} including stellar feedback only implicitly. Other high-resolution galaxy-scale simulations of MW-like galaxies found that (clustered) SNe exploding in dense
environments alone are capable of regulating global star formation as well as producing
winds that are highly mass loaded \citep{Martizzi+2019, Tress+2020}. Those
simulations similarly adopted a more explicit description of ISM physics, star
formation, as well as stellar feedback processes. Due to differences in the setup the winds
generated in the simulations of \cite{Martizzi+2019} have, however, slightly
higher mass loading factor ($0.5 \le \eta \le 50$) than in our simulations ($1
\le \eta \le 10$). Similar to our findings, in their simulations, using
only self-gravity and SNe feedback as well, the gas forms cold dense star-forming
clouds within which the stars that explode in SNe are born. These dense and cold clouds are generally not disrupted by the SNe \citep[][deduced from Figure 10 of the latter]{Martizzi+2019,Tress+2020}. 

Additionally, we explore in detail the ability of SNe in regulating star formation at different surface densities and find that SNe
alone are efficient enough to regulate star formation at solar neighborhood surface
densities and lower (\sigmaGas $\lesssim $ 20~\Msun/pc$^2$).
We find that the galaxy reaches, at radii similar to that of the solar
neighborhood ($r\simeq7-9~$kpc), a self-regulated steady-state.  This analysis depends on
different surface densities and can be compared to detailed simulations that model a portion of
the ISM in representative pieces of isolated, stratified, galactic discs with
solar neighbourhood-like properties using (magneto-)hydrodynamic (MHD)
simulations. Such simulations have found similar results where SNe feedback
alone was strong enough to regulate star formation and the vertical disc
structure \citep{Korpi+1999,Joung+MacLow2006, deAvillez+Breitschwerdt2004,
Henley+2015, Girichidis+2016}.
 
For higher surface density environments we find that SNe are not efficient enough
in regulating star formation and that gravity and shear are important drivers of the 
evolution within these surface densities. \citet{Li+2005, Li+2006} also report that
their results suggest that the non-linear development of gravitational instabilities
determine their global and local KS relation. We argue that additional feedback
processes, such as ionising radiation and stellar winds, would be needed in order to
further reduce the rate at which stars are formed. Detailed studies of the ISM
have shown that ionising radiation impacts the ambient ISM structure by heating
the dense gas phase in star-forming regions, injecting momentum into the ISM
and driving local turbulence. This has an effect on gas accretion onto the star
clusters and thus further regulates the local efficiency of star formation
\citep{Peters+2008, Gritschneder+2009, Geen+2015b, Peters+2017, Rathjen+2021}.
Similar effects have been shown for stellar winds \citep{Gatto2017}, although
the relative effect compared to ionising radiation may, in high density
regions, be lower \citep{Haid+2016, Geen+2021}\footnote{However, this finding
may be also subject to resolution effects.}. 

Galaxy-scale simulations of MW-like galaxies often try to include such
additional feedback processes from massive stars using a sub-grid model
approach \citep{Agertz+2013,Hopkins+2018,Marinacci+2019}.  Their simulations
include, on top of a SNe feedback model, many additional stellar feedback
channels, such as radiation from stars, stellar winds, as well as
photo-electric heating in \citep{Hopkins+2018}. They find that the added
stellar feedback channels help to regulate and suppress star formation and in
some cases help to increase galactic outflows. Interestingly, \cite{Agertz+2013}
find for their SNe-only simulation a very similar SFR to the one in \satinI. 
The sub-grid models used in those simulations rely, however, on a number of
assumptions regarding, for instance, the coupling between the radiation and the
gas such as the absorption of photons, mean free paths, optical depths, and
shielding.  

\citealp{Rosdahl+2015} avoid certain assumptions made in the simulations using
a sub-grid model for radiation by tracking the radiation from the stars directly using
radiation-hydrodynamics (RHD). They find that for a MW-like galaxy radiation
from stars helps to suppress star formation as much as the inclusion of SNe
feedback does. This is mainly due to the suppression of dense
cloud formation, rather than their destruction due to the radiation.  The importance
of radiation feedback (mainly photo-heating) to suppress star formation
decreases with galaxy mass in their simulations. In future work, it will be interesting to compare our approach following the stellar evolution of individual massive
stars and using realistic delay time distributions within the star cluster with the one of \cite{Rosdahl+2015} assuming an average stellar population. 

We have shown that in \satinI{} SNe launch outflows with mass loading factors
comparable with similar high-resolution MW-like simulations such as reported in
\cite{Martizzi+2019}, as well as \cite{Marinacci+2019} when they include all
their feedback channels\footnote{Note that they report lower SFR as well as
lower mass outflow rates leading, however, to a similar mass loading factor.}.
Additionally looking at the different gas phases in the outflow we have shown
that the majority of the gas that leaves the galaxy is hot, whereas the warm
and cold gas falls back onto the galaxy. Such a galactic fountain flow that
helps to sustain late-time star formation has also been measured in
\cite{Martizzi+2019} and \cite{Marinacci+2019} where most of their ejected gas
eventually rains back onto the galactic disc. 

At the same time, we do not achieve the strong outflows that are needed in
cosmological simulations in order to produce reasonable galaxies in the
$\Lambda$CDM cosmology \citep[e.g][]{Schaye+2015, Muratov+2015, Pillepich+2018, Tollet+2019}. 
Future simulations including additional
stellar feedback processes will be needed to assess whether such channels help
to increase the mass outflows in our simulations. Additionally, cosmological
simulations would be helpful to explore the potential of the \satin{}
model to produce realistic galaxies in a fully cosmological
context\footnote{Given the current implementation such a simulation, while
maintaining a good numerical resolution, would be computationally extremely
expensive if not impossible.}.  

Earlier work by \cite{Gatto2017} and \cite{Peters+2017} has shown that a VFF of
at least 50\% is needed in the hot phase to drive a galactic
outflow\footnote{This assumption has to be modified if cosmic rays are
considered in the simulations (see
\citealp[e.g.,][]{Peters+2015,Girichidis+2016, Simpson+2016}).}. This is in line
with a measured VFF of around 50\% in the hot phase in \satinI. 
Following is the VFF of the warm ($\sim$~30-35\%) and then the cold
phase ($\sim$~15-20\%). These obtained results can be compared to models for the ISM phase VFF that are
based on observations of the MW such as those shown in
\cite{Kalberla+Kerp2009}. 
Assuming turbulent pressure equilibrium, they derived
for the inner 250~pc a VFF in the cold phase of 5-18\%, in the warm phase
60-67\% (adding the contribution of their warm and warm-hot phase together),
and in the hot phase 17-23\% (see their Figure~11). Our results for the \SNe{}
simulation match the VFF of the cold phase, but underestimate/overestimate the VFF in the
warm/hot phase. The SNe-only stratified disc simulation by \cite{Peters+2017}
obtained similar results to our simulations, whereas only their simulation 
including radiation from
the stars approached observed values. This is because radiation disrupts the dense gas in
the star-forming regions and increases/decreases the VFF in the warm/hot phase. 

By changing the ISM structure, radiation likely also has an effect on the
energetics of the ISM. In our simulations, SN feedback alone is not strong
enough to disrupt the dense and cold clouds (similar to
\citealp{Rathjen+2021}). Looking at the power spectrum of the ISM we find that
most of the kinetic power of the cold gas is `locked-up' in the clumps and that
SNe alone are not capable to destroy the dense gas clumps and to drive turbulence 
in the cold gas on galactic scales.  In future work we plan to study the effect of radiation on the energetics of the ISM. Along similar lines, adding other early feedback channels such
as stellar winds and cosmic ray heating could help to
prevent the formation of the coldest gas within the galaxy, and likely have
an effect on the energetics of the ISM as well. Getting the VFF of the warm/hot phase correct and including other early feedback processes may also be important for the clustering of SF. \citet{Hislop+2022} get, when including only SNe feedback, a clustering that is too high and, similar to our simulation, high 
density clumps that are never destroyed.

Finally, we note that we do not only lack early stellar feedback
channels, such as radiation and winds, in our simulations but also 
have not taken into account other important physical processes in our model. Such processes include cosmic ray transport and physics, different cooling and heating processes (such as molecular cooling, cosmic ray and photo-electric heating), magnetic fields, dust production, destruction and evolution (both important for the molecular
chemistry as well as for reprocessing radiation fields), and thermal conduction.  
The inclusion of some of these processes will be part of future work. 

In summary, our general results from the \satinI{} simulations are similar to those found in other full MW-like galaxy
scale simulations with different implementations of star formation and SNe feedback and similar or coarser resolution. Additionally, the \satinI{} simulations also agree well with detailed simulations that model only portions of the ISM and only include SNe. Our investigation of star formation within different surface densities of the disc as well as comparing the ISM properties with observations suggests that other feedback channels may be needed in order to better match observed values, especially in higher surface density environments. Detailed ISM simulations have suggested similar in the literature. It will be interesting to investigate this within our \satin{} model and additional full galaxy-scale simulations in the future. 

\section{Conclusions}\label{sec:conclusion} 
In this paper we introduce the star formation and supernova (SN) feedback model of the
\satin{} project. The goal of this project is to investigate the interaction between 
AGN and a turbulent multi-phase ISM.  We model the formation of star
clusters with sink particles and track the evolutionary state of individual massive
stars that form within each cluster in order to get realistic SNe delay
times. The employed SN model adapts to the local environment depending on
whether the cooling radius of the Sedov blast wave is resolved. The \satin{} model
is a galaxy wide implementation of a successful ISM model and naturally covers an
order of magnitude in gas surface density, shear and radial motions. The model is 
adapted in the AMR code \Ramses{}. We test the implemented model in high-resolution isolated simulations of an isolated MW-like disc galaxy with a peak 
resolution of 9~pc. We find that
\begin{itemize}
\item  SNe alone are able to regulate star formation locally at 
solar neighborhood surface densities and lower. The galaxy establishes, because
of the feedback, a Kennicutt-Schmidt relation that matches globally and
locally with observations. 
\item Due to SNe feedback, the simulations capture a distinct multi-phase ISM with three phases (cold, warm, hot)
coexisting and interacting with each other.  With SNe, the gas in the ISM covers
larger regions in the phase diagram that are thermally unstable. The volume filling fraction of the cold
phase matches with observations, whereas the VFF of the warm/hot phase are underestimated/overestimated.
\item SNe are an important driver of the density and energy structure of the ISM. 
SNe drive additional turbulence in the warm gas and increase the kinetic energy 
of the cold gas, partly cooling out of the warm phase. SN feedback is not strong enough to disrupt dense and cold clouds and
to drive turbulence for the cold gas on galactic scales. At the same time
gravity, shear, and accretion also contribute significantly to the energetic structure of the ISM. 
\item SN feedback launches outflows with mass loading factors of $3 \le \eta \le 10$.
The majority of the gas that leaves the galaxy is hot. The warm and cold gas
mostly falls back onto the galaxy in a galactic fountain flow.
\end{itemize}
The inclusion of other stellar feedback processes from massive stars such as
radiation and stellar winds seems to be needed in order to reduce the rate
at which stars form in higher surface density environments, to
increase/decrease the VFF in the warm/hot phase, and to prevent the formation
of the very cold and dense clouds that we find in the simulations.
This will be subject of future work (Bieri et. al in prep.).
%
\section*{Acknowledgments}
We thank Freeke van de Voort and Francesca Fragkoudi for helpful comments
on the manuscript as well as interesting discussions. TN acknowledges support from the Deutsche Forschungsgemeinschaft (DFG, German Research Foundation) under Germany’s Excellence Strategy - EXC-2094 - 390783311 from the DFG Cluster of Excellence "ORIGINS". SG acknowledges support from a NOVA grant for the theory of massive star formation. During the time of his contribution to this work, JPC was employed at the Technical University of Munich.
%
\section*{Data Availability}
The simulation data used within this paper will be shared on reasonable request to the corresponding author.

\bibliographystyle{mnras}
\bibliography{refs}

\newpage
\appendix

\section{Effect of Resolution on Star Formation}
\label{sec:EffectOfResolutionOnStarFormation}
\begin{figure*}
\plottwoH{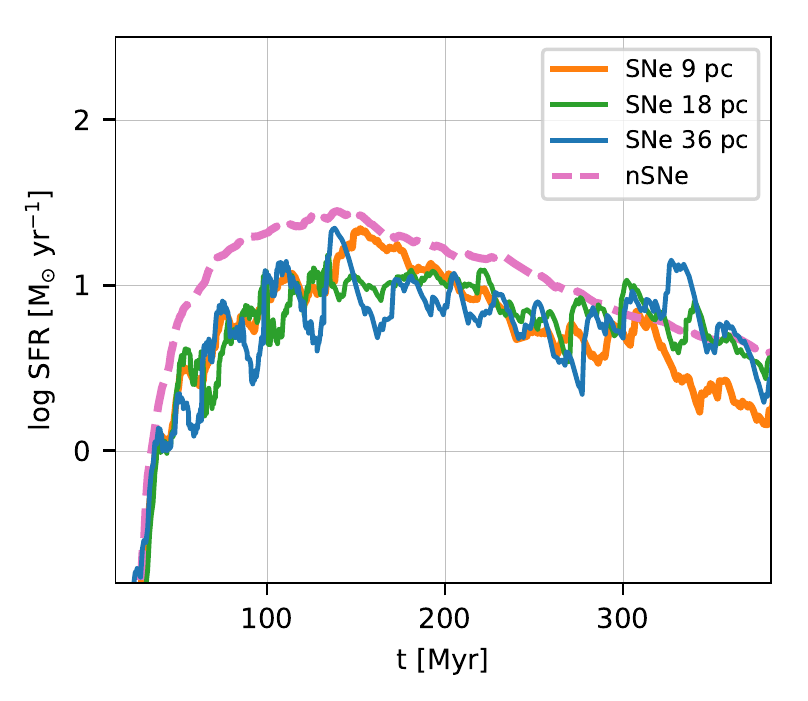}{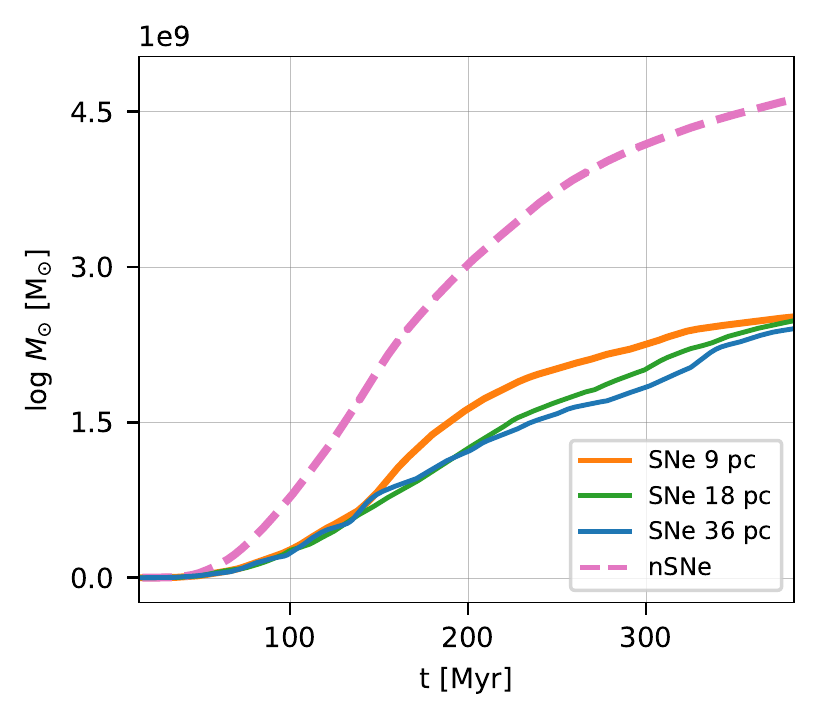}
\caption{Left: Global observed SFR history (see Eq.~\ref{eq:SFR_OB}) for SNe simulations with different
resolution (9~pc as standard, 18~pc, and 36~pc) in comparison with the 
\nSNe{} simulation. The dashed pink and orange line are the same as in Figure~\ref{fig:totalSFRtime}. The global SFR of the lower resolution simulations agree, especially at the beginning, quite well with the default \SNe{} simulation. When including SNe feedback the SFR oscillates more the coarser the resolution. The SFR of the \nSNe{} simulation is noticeably smoother than the SNe simulations. After an initial
rise the SFR decrease for all the simulations but much more stochastically when SNe is included. Towards the end of the simulations resolution has a comparable relative effect on the SFR as the SNe explosions. \newline
Right: The fact that the different resolutions agree well with each other is also clear from the total integrated stellar mass formed. All the different resolution simulations converge towards a similar value. The \nSNe{} simulation transforms much more gas into stars, as already discussed in the main text above (see Section \ref{sec:SFR-KS}).}
\label{fig:SFRResolution}
\end{figure*}
In order to test the effect of resolution on the formation of stars in 
our galaxy we have performed two other simulations, including SNe feedback, with progressively lower resolution: 18~pc ($=2 \times$ standard resolution) and 36~pc ($=4 \times$ standard resolution). In addition to the lower cell resolution we also changed the refinement criteria where in the lower resolution simulations a cell is refined if the gas within a cell is larger than $5 \times 10^4$~\Msun and $5 \times 10^5$~\Msun, respectively, instead of $5 \times 10^3$~\Msun as for the standard resolution. Everything else is kept the same between the different simulations (i.e., initial condition, number of stellar and DM particles, sink formation threshold) in order to best test the effect of gas resolution on the SFR within the galaxy. Figure~\ref{fig:SFRResolution} shows in the left panel the global observed SFR history for the different resolution simulations (including the default simulation \SNe{}) as well as the \nSNe{} simulation. The different resolution simulations agree quite well with each other. This is again confirmed in the 
right panel showing the total integrated stellar mass formed. Here the value for all the SNe 
simulations at different resolutions converge towards a similar value. The decrease in SFR for all the simulations is more stochastic when SNe is included and stronger with lower resolution. This could be because in lower resolution simulations the SNe manage to disrupt or destroy the high density gas more effectively causing the SFR to drop momentarily, only to increase when gas starts to collapse towards the high density clouds and stars start to form again. This also causes the total amount of formed stars to be slightly lower the coarser the resolution. Such complexities indicate that convergence (in this context convergence of star formation and the properties of the ISM within the galaxy) continues to pose a challenge for galaxy formation models.
\section{Computing Power Spectra}
\label{sec:PowerSpectrumCalculations}
We quantify the kinetic state of the ISM in Fourier space by first calculating the kinetic energy
power spectra and then divide it by the mass within the box to get the specific
kinetic energy power spectra (see Section~\ref{sec:ISM}). This improves the
comparison between simulations. The kinetic energy power spectra of 
the weight $w$ is defined as  
\begin{equation}
P \left( \mathbf{k} \right) = \tilde{w} \left( \mathbf{k} \right) \cdot w \left( \mathbf{k} \right) ^* \quad , 
\end{equation} 
where $\tilde{w} \left( \mathbf{k} \right)$ is the Fourier transform of the real array $w \left( \mathbf{k} \right)$, $\mathbf{k}$ is the wave vector, and `*' refers to the 
complex-conjugate. To get the Fourier transform of the kinetic energy field we used $w = \sqrt{\rho} \mathbf{v} \left( \mathbf{r} \right) $. 
Here $\mathbf{v} \left( \mathbf{r} \right)$ is the three dimensional velocity vector at
cell center $\mathbf{r}$ (measured in our case from the center of the galaxy) and $\rho$ the total density within each simulation cell. With this choice of weights, the
power spectrum directly measures the kinetic energy (see
also \citealp{Grisdale+2017}). Another weight that is often used
($w = \rho ^{1/3} \mathbf{v}$) corresponds to a kinetic energy flux and has been
shown to reproduce energy spectra with Kolmogorov scaling ($e_{\mathbf{k}}
\propto \mathbf{k}^{-5/3}$) in super-sonic flows \citealp[see][for this choice]{Kritsuk+2007,Kowal+2007, Schmidt+2008, Federrath+2009}.

We calculate the power spectrum within a $\sim 10\times10\times10$~kpc$^3$
region centred on the galaxy\footnote{More precisely, we calculate the size of the
box for the power spectrum such that a power of two number of cells covers the region using the given cell resolution.}. 
For all of the spectra we use a uniform grid of resolution $\sim$~40~pc and
account for the non-periodic boundary by zero-padding the calculated cubes. The
length of the void domain is in each dimension set to be twice as big as the
chosen box size. We experimented with more padding as well as different box
sizes and confirmed that the main results and findings are not altered. 

To obtain the power spectrum $P (\mathbf{k})$ of the specific kinetic energy we first Fourier
transform the cube, bin $P (\mathbf{k})$ in wave vectors $ k = \left| \mathbf{k}
\right|$, normalise by the area of the spherical surface with radius k + 0.5, and
finally divide it by the total mass within the box. This finally gives the
`angle-averaged' specific kinetic energy spectrum \citealp[e.g.][]{Joung+2006}.
$\left< P(k) \right>$, where the physical scale $l$ is connected to the wave
vector via $k = 2 \pi / l$.  

 \bsp	
 \label{lastpage}
 \end{document}